# Generalized Rational Variable Projection With Application in ECG Compression

Péter Kovács, Sándor Fridli, and Ferenc Schipp

*Abstract*—In this paper we develop an adaptive transform-domain technique based on rational function systems. It is of general importance in several areas of signal theory, including filter design, transfer function approximation, system identification, control theory etc. The construction of the proposed method is discussed in the framework of a general mathematical model called variable projection. First we generalize this method by adding dimension type free parameters. Then we deal with the optimization problem of the free parameters. To this order, based on the well-known particle swarm optimization (PSO) algorithm, we develop the multi-dimensional hyperbolic PSO algorithm. It is designed especially for the rational transforms in question. As a result, the system along with its dimension is dynamically optimized during the process. The main motivation was to increase the adaptivity while keeping the computational complexity manageable. We note that the proposed method is of general nature. As a case study the problem of electrocardiogram (ECG) signal compression is discussed. By means of comparison tests performed on the PhysioNet MIT-BIH Arrhythmia database we demonstrate that our method outperforms other transformation techniques.

## I. Introduction

ANALYSIS of signals by means of mathematical transformations proved to be an effective method in various aspects. For instance, dimensionality reduction methods are strongly related to the problems of compression and noise suppression of the original signal. Moreover, the transform can also be used for extracting features in classification tasks. Many of these transform-domain techniques are generated by fixed basic functions like the trigonometric functions in the Fourier transform, Walsh functions in the Walsh–Fourier transform, mother wavelet function for the wavelet transform, etc. In order to surpass the limitations in compression ratio, reconstruction error, adaptivity and computational complexity of these algorithms, the dictionary based methods such as matching and basis pursuit were proposed by many authors, see e.g. [1], [2]. In that case, an overcomplete set of base functions was applied to increase the adaptivity. Also, the wavelet packet transform (WPT), which utilizes different tilings of the time-frequency plane, was introduced by Coifman *et al.* [3]. Although these led to improved adaptivity, the performance remained limited because of the lack of free parameters.

There is a trade-off between overfitting and underfitting, which is controlled by the number of parameters, i.e., the model order. Information theory provides many order selection rules, which quantify the optimized models according to the Bayesian information criterion (BIC), Akaike information criterion (AIC), etc. In this case, the BIC and/or the AIC of the candidate models are precalculated for various parameter setups, then the best among them is selected according to the given criterion [4]. Basis pursuit can also be applied to make an initial assumption on the number of basic function, which is followed by an additional optimization step to find the best nonlinear parameters.

The orthogonal systems of rational functions play distinguished role and proved to be very effective in several areas and applications. Although their properties, like simplicity and the high variety of such systems, make them promising candidates for adapting transform domain technique there are only few examples for their application in signal processing. Moreover, the methods used in those examples do not employ the capacity of such systems completely.

In this paper we systematically develop a new method that fully utilizes the versatility of rational systems. To this end, we start from the well-known variable projection functional [5] and generalize it by introducing a new cost function. In the generalized variable projection method we integrate system dimensionality into the set of free parameters. In contrast to previous approaches, we jointly optimize both the accuracy and the complexity of the model. By means of rational systems we demonstrate that the increased computational demand is manageable, and the generalized method is efficient. Toward this we set up a new architecture space that supports the structure of the parameter space (number of poles, multiplicities) of the rational functions. We develop the hyperbolic version, based on the Poincaré model, of the stochastic Multi-Dimensional Particle Swarm Optimization (MDPSO) [6], [7] for the nonlinear optimization problem of system parameters. Finally, ECG compression with the analysis of the parameters and

Manuscript received December 21, 2018; revised August 6, 2019 and December 4, 2019; accepted December 11, 2019. Date of publication December 20, 2019; date of current version January 20, 2020. The associate editor coordinating the review of this manuscript and approving it for publication was Dr. Yuichi Tanaka. This work was supported in part by the Hungarian Scientific Research Funds (OTKA) under Grant K115804, in part by EFOP-3.6.3-VEKOP-16-2017-00001: Talent Management in Autonomous Vehicle Control Technologies, which was supported by the Hungarian Government and co-financed by the European Social Fund, and in part by ÚNKP-17-4 New National Excellence Program of the Ministry of Human Capacities. *(Corresponding author: Péter Kovács.)*

The authors are with the Department of Numerical Analysis, Faculty of Informatics, Eötvös Loránd University, Budapest 1117, Hungary (e-mail: kovika@inf.elte.hu; fridli@inf.elte.hu; schipp.ferenc@gmail.com).

This article has supplementary downloadable material available at http://ieeexplore.ieee.org, provided by the authors.

Digital Object Identifier 10.1109/TSP.2019.2961234

comparisons with the state-of-the-art methods is taken as case-study. The comparison tests show that our method outperforms other transformation algorithms.

Regarding the application, the main advantage of the proposed method is its adaptivity. State-of-the-art ECG processing methods usually fix the function system a priori based on the shape similarity between the ECGs and the basic functions. For this reason, Hermite functions, B-splines, and wavelets became popular in this field. Although the shape of the basic functions correlates very well with the normal ECG morphology, it is difficult to represent abnormal beat classes. The proposed method automatically scales up the number of free parameters in order to approximate abnormal beats with an acceptable level of error.

The paper is organized as follows. Section II contains the projection methods and the corresponding function systems. In Section III we develop the generalized variable projection model. In order to solve the corresponding optimization problem we extend the basic and multi-dimensional PSO algorithm using the Poincaré model of the hyperbolic geometry in Sections IV–V. In Section VI we apply our technique for the construction of an ECG compression method using optimized rational functions. In Section VII a comparative study of different adaptive transform-domain based techniques evaluated on the MIT-BIH arrhythmia database [8] is provided. The discussion of the results can be found in Section VIII. Finally, Section IX is a summary of conclusions and future plans.

## II. Projection Methods

This section serves as background and framework, along with examples, for the construction of generalized variable projection method presented in Section III.

### A. Non-Variable Projection Method

Let us start with the classical non-variable projection method. By projections we will always mean orthogonal projection, which is an important transform-domain technique. It is strongly related to approximation theory in Hilbert spaces. Namely, the space of the appropriate signals is usually considered to be a Hilbert space $\mathcal{H}$ with scalar product $\langle \cdot, \cdot \rangle$. Then, a signal $f \in \mathcal{H}$ is modeled by its orthogonal projection onto a closed subspace $\mathcal{S} \subset \mathcal{H}$. In practical applications $\mathcal{S}$ is a finite $N \in \mathbb{N}_+$ dimensional subspace spanned by a linearly independent function system $\Theta := \{ \Theta_j \in \mathcal{H} : 0 \leq j < N \}$. It is well-known that for any signal $f \in \mathcal{H}$ the best approximation $\widetilde{f} \in \mathcal{S}$ uniquely exists: $\|f - \widetilde{f}\| = \inf \{ \|f - g\| : g \in \mathcal{S} \}$ and $f - \widetilde{f} \perp \mathcal{S}$. The norm $\|f\| = \sqrt{\langle f, f \rangle}$ is the usual one induced by the scalar product. The orthogonal projection from $\mathcal{H}$ onto $\mathcal{S}$, which is a continuous linear operator, will be denoted by $P_\Theta^N$. Then, a signal $f \in \mathcal{H}$ is represented in $\mathcal{S}$ as follows:

$$\widetilde{f}(t) = P_\Theta^N f(t) := \sum_{j=0}^{N-1} c_j \Theta_j(t) \quad (t \in \Omega), \qquad (1)$$

where the coefficients $c_j \in \mathbb{R}$ $(j = 0, \ldots, N-1)$ are the solution of the system of linear equations $\mathbf{Gc} = \mathbf{b}$. In this equation $\mathbf{G} = \big( \langle \Theta_i, \Theta_j \rangle \big)_{i,j=0}^{N-1}$ denotes the usual Gram matrix, $\mathbf{c} = (c_0, \ldots, c_{N-1})^T$, and $\mathbf{b} = \big( \langle f, \Theta_0 \rangle, \ldots, \langle f, \Theta_{N-1} \rangle \big)^T$. If $\{ \Theta_j(\cdot) : 0 \leq j < N \}$ is a set of orthonormal functions, then $\mathbf{G}$ is the identity matrix and $c_i = \langle f, \Theta_i \rangle$ is the $i$th Fourier coefficient of $f$. A typical example for the space of appropriate real valued signals is $\mathcal{H} = L_w^2(\Omega)$ with a positive weight function $w$. $\Omega$ is usually an interval, bounded or unbounded, for analog signals and a proper countable set of real numbers for discrete-time signals. Various types of basic functions have been used so far depending on the particular problem. For instance, trigonometric functions are applied in MP3 coding, while wavelets in JPEG 2000 standards. In these cases the shapes of the basic functions are fixed, they cannot be adjusted to the individual signal. This limitation proved to be important, especially in dynamically changing environments, where we need to adjust the system to the signal. For instance, in case of ECGs, normal beats usually dominate the signal, and even the non-adaptive techniques work well on them. On the other hand, abnormal beats can have rather complex waveforms for which those techniques provide poor results. These phenomena raise the need for more sophisticated models, in which the basic functions can be adapted to the shape of the signal due to their free parameters.

### B. Variable Projection Method

The theory of variable projection methods was laid down by Golub and Pereyra in [5]. They provided formulas for theoretical and numerical derivation for the related Gauss–Newton type algorithms. Since then these methods have found applications in many areas including neural networks, telecommunications, dynamical systems, etc. A good summary on them are given in [9], [10]. We point out that several well-known transformations, e.g. B-splines, orthogonal polynomials, can be understood as variable projections. As a consequence the related results can be interpreted in a unified framework, which has avoided the attention so far. The above-mentioned special models are discussed in the next subsection.

Let us now suppose that instead of a fixed function system we have a collection of systems $\{ \Theta(\mathbf{a}) : \mathbf{a} \in \Gamma_1 \}$ where $\Gamma_1$ is an index set. We will assume that for every $\mathbf{a} \in \Gamma_1$ the function system $\Theta(\mathbf{a})$ is of the form $\Theta(\mathbf{a}) := \{ \Theta_j(\cdot, \mathbf{a}) \in \mathcal{H} : 0 \leq j < N \}$, where $N \in \mathbb{N}$ is fixed and the $\Theta_j(\cdot, \mathbf{a})$'s are linearly independent. Then the general nonlinear model is

$$\eta(\mathbf{c}, \mathbf{a}; t) := \sum_{j=0}^{N-1} c_j \Theta_j(t, \mathbf{a}) \quad (t \in \Omega, \ \mathbf{a} \in \Gamma_1), \qquad (2)$$

where the $c_j$'s are real or complex coefficients. In order to get the best model for a particular analog signal $f \in \mathcal{H}$, we need to find the optimal values for $\mathbf{c}$ and $\mathbf{a}$, i.e., to minimize the nonlinear functional

$$r(\mathbf{c}, \mathbf{a}) := \|f(\cdot) - \eta(\mathbf{c}, \mathbf{a}; \cdot)\|. \qquad (3)$$

Let us denote the $\mathbf{a}$ dependent Gram matrix by $\mathbf{G}(\mathbf{a})$. For a fixed parameter $\mathbf{a}$ the optimization with respect to $\mathbf{c}$ is linear and the result is the orthogonal projection denoted by $P_{\Theta(\mathbf{a})}^N f$. We note that $\mathbf{c}(\mathbf{a})$ can be calculated by solving $\mathbf{G}(\mathbf{a})\mathbf{c}(\mathbf{a}) = \mathbf{b}(\mathbf{a})$.

Using this fact, the original problem in Eq. (3) can be reduced to minimize

$$r_2(\mathbf{a}) := r(\mathbf{c}(\mathbf{a}), \mathbf{a}) = \left\| f(\cdot) - P^N_{\Theta(\mathbf{a})} f(\cdot) \right\|, \quad (4)$$

which following Golub and Pereyra [5] is called *variable pojection functional*.

In applications, we work with discrete signals $\mathbf{f} \in \mathbb{R}^M$ and with matrices $\Theta(\mathbf{a}) \in \mathbb{R}^{M \times N}$ representing discrete function systems. The formulas developed above for the analog case can easily be adjusted to obtain their discrete versions. Namely, the coefficient vector $\mathbf{c}(\mathbf{a}) \in \mathbb{R}^N$ can be calculated via $\mathbf{c}(\mathbf{a}) = \Theta(\mathbf{a})^+ \mathbf{f}$, where $\Theta(\mathbf{a})^+$ is the Moore–Penrose generalized inverse of $\Theta(\mathbf{a})$. Moreover, the discrete variant of the orthogonal variable projection operator is $\mathbf{P}^N_{\Theta(\mathbf{a})} = \Theta(\mathbf{a})\Theta(\mathbf{a})^+$. Then the discrete variable projection functional is of the form

$$r_2(\mathbf{a}) = \left\| \mathbf{f} - \mathbf{P}^N_{\Theta(\mathbf{a})} \mathbf{f} \right\|_2 = \left\| \mathbf{f} - \Theta(\mathbf{a})\Theta(\mathbf{a})^+ \mathbf{f} \right\|_2, \quad (5)$$

where $\| \cdot \|_2$ stands for the usual Euclidean norm in $\mathbb{R}^M$. In [5], Golub and Pereyra showed that the functionals $r(\mathbf{c}, \mathbf{a})$ and $r_2(\mathbf{a})$ have the same global minima in both the analog and the discrete versions. Furthermore, they demonstrated the fact that iterative nonlinear algorithms converge faster on the reduced $r_2(\mathbf{a})$ problem.

### C. Examples for Variable Projections

Now we show examples that can be understood as special variable projections. Consequently the optimization problems involving them can be reduced to the form given in Eq. (5). Although the theoretical framework is the same in the discussed cases the corresponding algorithms can be different in several respects, such as efficiency, complexity. We demonstrate the pros and cons using the example of ECG compression.

*1) B-Splines:* We consider the optimization of B-splines $B_{j,\ell}(\cdot, \mathbf{a})$ ($n \in \mathbb{N}^+$, $j = 0, \ldots, n-1$, $\mathbf{a} \in \mathbb{R}^n$) of degree $\ell \in \mathbb{N}$, which are defined by

$$B_{j,0}(t, \mathbf{a}) := \begin{cases} 1 & \text{if } a_j \leq t < a_{j+1}, \\ 0 & \text{otherwise}, \end{cases}$$

$$B_{j,\ell}(t, \mathbf{a}) := \beta_{j,\ell} B_{j,\ell-1}(t, \mathbf{a}) + \gamma_{j,\ell} B_{j+1,\ell-1}(t, \mathbf{a}),$$

where

$$\beta_{j,\ell} = \frac{t - a_j}{a_{j+\ell} - a_j}, \quad \gamma_{k,\ell} = \frac{a_{j+\ell+1} - t}{a_{j+\ell+1} - a_{j+1}}.$$

Let us take the B-splines with fixed degree $\ell$. Then the base functions in the variable projection model are $\Theta_j(t, \mathbf{a}) = B_{j,\ell}(t, \mathbf{a})$, and the free parameter is the vector of free knots $\mathbf{a} = (a_0, a_1, \ldots, a_{n-1})^T$. The problem is to find the optimal locations of the knots. Unfortunately, the reduced functional $r_2(\mathbf{a})$ has several stationary points in this case, which makes the optimization quite a hard task. This phenomenon, called *lethargy problem*, was thoroughly studied by Jupp in [11]. We note that the B-spline model with free knots was adapted to ECG data compression tasks by Karczewicz and Gabbouj in [12]. Here, the optimization process is an iterative method, which removes the least significant knot at each step. Because of its high flexibility both the approximation and compression properties of the spline algorithm are comparable with the recent methods (see e.g. Table IV). On the other hand, the computational cost is high due to lack of orthogonality.

*2) Hermite Orthogonal Polynomials:* Orthogonal polynomials are widely used in signal processing. In particular, Hermite polynomials have found many applications in ECG compression [13], [14], classification [15], [16], feature analysis [17] and QRS detection [18]. Classical Hermite polynomials are defined by the following recursion:

$$H_{j+1}(t) = 2tH_j(t) - 2jH_{j-1}(t) \quad (t \in \mathbb{R}, \ j \in \mathbb{N}^+),$$

where $H_0(t) = 1$ and $H_1(t) = 2t$. The so-called *Hermite functions* are constructed from them by dilation:

$$\varphi_j(t, a) = \frac{1}{\sqrt{a 2^j j! \sqrt{\pi}}} e^{-\frac{t^2}{2a^2}} H_j\left(\frac{t}{a}\right) \quad (t, a \in \mathbb{R}, \ a > 0).$$

These functions form a complete orthonormal system in $L^2(\mathbb{R})$. In the Hermite type model of ECG signals, three individual variable projection methods are combined. Each of them is based on a dilated Hermite system $\Theta_j(t, a_i) := \varphi_j(t, a_i)$ ($0 \leq i < 3$) and is used to represent the corresponding segments P, QRS, T of the heart beat. The best value of each dilation parameter $a_i$ is determined via optimization. It is worth mentioning that in a recent paper [14] on discrete Hermite functions a significant improvement is presented in terms of compression of the QRS complex. In Section VII we enhanced the original algorithm in [13] by combining it with [14] and used it in our comparative study. For the theory of orthogonal polynomials and numerical algorithms we refer to the fundamental books [19], [20].

*3) Wavelets:* The discrete wavelet transform (DWT) is one of the most popular transforms in signal processing [21], [22]. The construction of wavelet transforms (WT) is based on a so-called *scaling function* $\phi$ for which the translates $\{\phi(t-k)\}_{k \in \mathbb{Z}}$ form an unconditional orthonormal basis in the initial subspace $V_0 \subset L^2(\mathbb{R})$. $V_0$ is supposed to be invariant with respect to integer translations. It generates the multiresolution $V_i \subset V_{i+1}$ ($i \in \mathbb{Z}$) with the condition $v(t) \in V_i$ if and only if $v(2t) \in V_{i+1}$. Additionally $\bigcup_{i=0}^\infty V_i$ must be dense in $L^2(\mathbb{R})$. Then, the original signal is decomposed according to the subspaces $V_i$ and $W_i$, where $W_i$ is the orthogonal complement of $V_i$ with respect to $V_{i+1}$. The corresponding bases in $V_i$ and $W_i$ are respectively

$$\phi_{i,k}(t) = 2^{i/2} \phi(2^i t - k) \quad (i, k \in \mathbb{Z}),$$

$$\psi_{i,k}(t) = 2^{i/2} \psi(2^i t - k) \quad (i, k \in \mathbb{Z}),$$

where $\psi$ is the so-called *mother wavelet* induced by $\phi$.

In practice we deal with discrete wavelet transform for finite signals, which can be viewed as periodic signals. There the scaling function $\phi$ and the *mother wavelet* $\psi$ are completely characterized by a compactly supported low-pass filter $\mathbf{h}$ as follows:

$$\phi(t) = \sum_{k=0}^{L-1} \mathbf{h}_k \phi(2t-k), \quad \psi(t) = \sum_{k=0}^{L-1} \mathbf{g}_k \phi(2t-k),$$

where $\mathbf{g}_k = (-1)^k \mathbf{h}_{L-1-k}$ is a high-pass filter and $L$ is the filter length. Although the constraints of orthogonality restrict the values of the filter's coefficients, we still have $\frac{L}{2} - 1$ degree-of-freedom to choose $\mathbf{h}_k$ (see e.g. Sect. 5.9 in [23]). We note that, for ECG modeling [24], [25] the filter dimension is usually $L = 6$. Then there are two free parameters $a_1$ and $a_2$ which determine the filter's coefficients $\mathbf{h}_k$ [23]:

$$k = 0, 1: \quad \mathbf{h}_k = \frac{1}{4\sqrt{2}} \cdot [(1 + (-1)^k \cos a_1 + \sin a_1)$$
$$\cdot (1 - (-1)^k \cos a_2 - \sin a_2)$$
$$+ (-1)^k 2 \sin a_2 \cdot \cos a_1],$$
$$k = 2, 3: \quad \mathbf{h}_k = \frac{1}{2\sqrt{2}} \cdot [(1 + \cos(a_1 - a_2)$$
$$+ (-1)^k \sin(a_1 - a_2))],$$
$$k = 4, 5: \quad \mathbf{h}_k = \frac{1}{\sqrt{2}} - \mathbf{h}_{i-4} - \mathbf{h}_{i-2} \quad (a_1, a_2 \in [0, 2\pi]). \tag{6}$$

Now we insert the DWT into our framework by defining the vector index $\mathbf{j} = (i, k)$. Then, for a fixed $L = 6$ dimension we have $\Theta_{\mathbf{j}}(t, \mathbf{a}) := \psi_{\mathbf{j}}(t, \mathbf{a})$, where $\mathbf{a} = (a_1, a_2)^T$.

*4) Rational Functions:* In our last example we consider rational function systems, which can be viewed as the generalization of polynomial systems [26]. To this order let $\mathbb{C}$ stand for the set of complex numbers, $\mathbb{D} := \{z \in \mathbb{C} : |z| < 1\}$ for the open unit disc, $\mathbb{T} := \{z \in \mathbb{C} : |z| = 1\}$ for the unit circle (or torus). For a sequence $\mathbf{a} = \{a_n \in \mathbb{D}\}_{n \in \mathbb{N}}$ the elements of the corresponding orthogonal system, which is called *Malmquist–Takenaka* (MT) system [27], [28], can be given in an explicit form as follows:

$$\Phi_j(z, \mathbf{a}) = \frac{\sqrt{1 - |a_j|^2}}{1 - \overline{a}_j z} \prod_{k=0}^{j-1} B(z, a_j) \quad (j \in \mathbb{N}),$$

where $B(z, a)$ is the so-called *Blaschke function* defined by

$$B(z, a) := \frac{z - a}{1 - \overline{a}z} \quad (z \in \mathbb{C} \setminus \{1/\overline{a}\}).$$

The parameter $a$ is called inverse pole, where $1/\overline{a}$ is the pole in the usual sense.

For the finite-dimensional version let $\mathbf{a} = (a_0, \ldots, a_{n-1})^T \in \mathbb{D}^n$ be a vector of distinct inverse poles with multiplicities $\mathbf{m} = (m_0, \ldots, m_{n-1})^T \in \mathbb{N}_+^n$. Then we will consider the MT system that corresponds to the inverse pole vector:

$$\mathbf{b} := (\underbrace{a_0, \ldots, a_0}_{m_0}, \ldots, \underbrace{a_{n-1}, \ldots, a_{n-1}}_{m_{n-1}})^T \in \mathbb{D}^N, \tag{7}$$

where $N = m_0 + m_1 + \ldots + m_{n-1}$. We note that although the MT system itself depends on the order of the inverse poles, the generated subspace and so the projection is invariant with respect to it. In this setting a signal $f$ belonging to the Hardy space $H^2(\mathbb{D})$ is modeled by taking $\Theta_j(t, \mathbf{b}) = \Phi_j(e^{it}, \mathbf{b})$ in Eq. (1) with

$$c_j = \langle f(\cdot), \Phi_j(\cdot, \mathbf{b}) \rangle = \frac{1}{2\pi} \int_{-\pi}^{\pi} f(e^{it}) \overline{\Phi}_j(e^{it}, \mathbf{b}) \, dt, \tag{8}$$

which are called the MT –Fourier coefficients. Applying appropriate discretization algorithms such as those in [29], [30], the integral above can be substituted by finite sums (see Theorem 2. in [31]). We note that this model was effectively used in QRS modeling [32], system identification [33], EEG seizure classification [34]–[36], sleep stage classification [37], etc.

### III. GENERALIZED VARIABLE PROJECTION METHOD

Let us start with the variable projection model and the corresponding functional given in Eq. (4). We note that in this model the dimension $N$ of the subspace is a priori fixed. Our aim is to develop a generalization by dropping this constraint, i.e., adding a new free parameter related to the dimension of the subspace. Toward the definition of the generalized variable projection method we define the new index set as $\Gamma \subset \Gamma_1 \times \Gamma_2$. In simple cases $\Gamma_2 = \mathbb{N}$ represents the dimension itself. This is the situation in Section II-C on Hermite functions. In that case increasing $N$ results in nested subspaces and so in a better minimum of the nonlinear functional Eq. (4). Consequently the optimization would terminate at the highest possible value of the dimension. On the other hand, high dimensions are not desired in real applications because it increases the complexity of the model. For controlling the dimension we introduce a penalty function $\Lambda(N)$ that is monotonically increasing. For the rest of the paper we always assume that $f$ is of zero mean with unit variance. Then the generalized variable projection functional including the penalty term is defined as follows:

$$\rho : \Gamma_1 \times \mathbb{N} \mapsto \mathbb{R}, \ \rho(\mathbf{a}, N) := \left\| f(\cdot) - P^N_{\Theta(\mathbf{a})} f(\cdot) \right\| + \Lambda(N). \tag{9}$$

There are however more complex cases when the parameters in $\Gamma_2$ are not simply dimensions, and the subspaces are not embedded into each other. In order to address this problem we modify the definition above to obtain the final form for the generalized variable projection functional

$$\rho : \Gamma \mapsto \mathbb{R}, \ \rho(\mathbf{a}, d) := \left\| f(\cdot) - P^{N(d)}_{\Theta(\mathbf{a}, d)} f(\cdot) \right\| + \Lambda(d), \tag{10}$$

where $\Theta(\mathbf{a}, d) := \{\Theta_j(\cdot, \mathbf{a}, d) \in \mathcal{H} : 0 \leq j < N(d)\}$ and $\Lambda(d)$ increasing in $d$ measures the complexity of the corresponding system in some sense. It may depend not only on the system but also on the specific task.

It is easy to see that the variable projection functional can be considered as a special case in which $\Gamma_2$ has exactly one element. We note that the main advantage of the generalized method is the simultaneous optimization with respect to the system and the dimension parameters. On the other hand, it makes sense only if all of the following conditions hold for the system:

1) it is flexible enough but easy to parametrize;

2) the complexity function is properly designed;

3) an efficient optimization can be constructed. (11)

We want to underline the fact that the rational functions are satisfy all these conditions. We will demonstrate the feasibility of rational function systems for generalized variable projection via signal compression problems. One of the key questions is to find an effective optimization algorithm. A generalized PSO type algorithm will serve our purpose. In the next section we establish the construction of it by starting from the basic version.

## IV. OPTIMIZATION OF RATIONAL VARIABLE PROJECTION BY PSO

### A. Basic PSO Algorithm

The basic PSO algorithm was introduced by Eberhart and Kennedy [38] as a population based stochastic optimization technique. In case of $n$ dimensional search space, the method is initialized by a random population $\{ \mathbf{x}_k \in \mathbb{R}^n : 1 \leq k \leq S \}$, where $S \in \mathbb{N}_+$ denotes the size of the swarm and every $\mathbf{x}_k$ is a potential solution for the optimization problem. The $\mathbf{x}_k$'s correspond to inverse pole configurations in our problem. For every $\mathbf{x}_k \in \mathbb{R}^n$ let $\widetilde{\mathbf{y}}_k \in \mathbb{R}^n$ denote the personal and let $\widehat{\mathbf{y}} \in \mathbb{R}^n$ denote the global best solutions achieved so far. In each step, both the position and the velocity of the particles are updated in the following way:

$$\mathbf{v}_k = c_1 r_1 \cdot (\widetilde{\mathbf{y}}_k - \mathbf{x}_k) + c_2 r_2 \cdot (\widehat{\mathbf{y}} - \mathbf{x}_k) + w \cdot \mathbf{v}_k,$$
$$\mathbf{x}_k = \mathbf{x}_k + \mathbf{v}_k \qquad (k = 1, \ldots, S), \qquad (12)$$

where the learning factors $c_1, c_2$ are predefined constants and $r_1, r_2 \in (0, 1)$ are uniformly distributed random numbers. The inertia weight $w$ was introduced later [39] in order to control the overall behavior of the swarm. For instance, one can favor exploration in the first few steps by increasing the value of $w$. Arbitrary large jumps are usually inhibited in the search space. To this end, the velocities and the positions are restricted to a certain interval defined by the parameters, $V_{\max}, X_{\min}, X_{\max}$. We note that following the standard we use this algorithm by setting $c_1 := 1.5, c_2 := 2$. Moreover, $w$ is linearly decreasing from 0.8 to 0.2. For other strategies of the parameter selection and convergence analysis we refer to [40], [41].

### B. Hyperbolic PSO Algorithm for Single-Pole Problems

In this section we develop the hyperbolic variant, inspired by single-pole rational optimization, of the PSO method. In this case the particles contain only two coordinates, i.e., the real and imaginary parts of the inverse pole. If the algorithm terminates in the $k$th optimal particle, then the optimal inverse pole is $\mathbf{x}_{k,1} + i\mathbf{x}_{k,2}$. Furthermore, as we know from Section II-C4, the inverse poles of the MT system must belong to the open unit disc $\mathbb{D}$. Hence, the search space is $\mathbb{D}$. This implies the idea to use the Poincaré model of the hyperbolic geometry to keep the particles within the search space. According to this idea, we will replace the arithmetic operators in Eq. (12) by their hyperbolic variants. We note that in this way the constrained optimization problem converts to non-constrained one. Moreover, by using proper mappings it can be applied to regions more general than the unit disc. [42] serves as a general reference work in this section.

*1) Hyperbolic Scaling:* Using the terminology of Euclidean geometry, the vector scalar multiplication of the hyperbolic space can be defined in a similar way. Namely, it means the scaling of a hyperbolic vector by keeping its direction. In this case, the geodesics of this space are represented by arcs of circles that are orthogonal to the torus. We recall the definition of the hyperbolic metric

$$\rho(z_1, z_2) := \operatorname{arctanh}(|B(z_2, z_1)|) \qquad (z_1, z_2 \in \mathbb{D}),$$

for which $(\mathbb{D}, \rho)$ is a complete metric space. This metric space is invariant with respect to the Blaschke transforms $\mathcal{B}(t, \mathfrak{a}) := \epsilon B(t, a)$, where $\mathfrak{a} := (a, \epsilon) \in \mathbb{D} \times \mathbb{T}$. We will use the fact that the hyperbolic segments can be defined via $\mathcal{B}(t, \mathfrak{a})$, which maps the interval $[0, p]$ onto the hyperbolic segment connecting $w_1$ and $w_2$, where

$$p = |B(w_2, w_1)|, \quad \epsilon = \frac{B(w_2, w_1)}{|B(w_2, w_1)|}, \quad a = -\bar{\epsilon} w_1. \quad (13)$$

Now the hyperbolic vector $\overrightarrow{w_1 w_2}$ can be defined as a directed segment with $\mathcal{B}(0, \mathfrak{a}) = w_1$ and $\mathcal{B}(p, \mathfrak{a}) = w_2$. Let us consider the scaling of a hyperbolic vector $\overrightarrow{w_1 w_2}$ by the factor $\lambda \in \mathbb{R}$. The new endpoint $w_\lambda$ of the solution vector $\overrightarrow{w_1 w_\lambda}$ is

$$w_\lambda := \epsilon B(s_{\lambda,a}), \quad s_\lambda = \tanh(\lambda \operatorname{arctanh}(p)). \quad (14)$$

In summary, the hyperbolic scaling $\lambda \odot \overrightarrow{w_1 w_2} := \overrightarrow{w_1 w_\lambda}$ can be evaluated with $w_\lambda = \mathcal{B}(s_\lambda, \mathfrak{a})$ as Eqs. (13)–(14) for any $\lambda \in \mathbb{R}$.

*2) Hyperbolic Addition:* It turned out that the right way to define the hyperbolic addition is based on the compositions of Blaschke functions. Namely, it can be shown that the collection of Blaschke transforms $\mathfrak{B} := \{\mathcal{B}(\cdot, \mathfrak{a}) : \mathfrak{a} \in \mathbb{D} \times \mathbb{T}\}$ is closed for composition. Moreover, $(\mathfrak{B}, \circ)$ is a subgroup of the well-known Möbius transformations, which maps the unit disc onto itself. In particular, if $\mathfrak{a}_1 = (w_1, 1)$ and $\mathfrak{a}_2 = (w_2, 1)$, then $\mathcal{B}(\cdot, \mathfrak{a}_1) \circ \mathcal{B}(\cdot, \mathfrak{a}_2) = \mathcal{B}(\cdot, \mathfrak{a})$, where $\mathfrak{a} = (w, \epsilon)$ with $\epsilon = \dfrac{1 + w_1 \overline{w_2}}{1 + \overline{w_1} w_2}$ and $w = \dfrac{w_1 + w_2}{1 + w_1 \overline{w_2}}$. The latter formula can be interpreted as a vector addition in the hyperbolic space for vectors with initial point at zero and endpoints at $w_1, w_2$ (see Sections 3.4-3.5 in [43]). Following this, we will use the operations

$$\overrightarrow{0 w_1} \oplus \overrightarrow{0 w_2} := \overrightarrow{0 w}, \quad \text{where} \quad w = \frac{w_1 + w_2}{1 + w_1 \overline{w_2}} \quad (w_1, w_2 \in \mathbb{D}),$$

$$\overrightarrow{0 w_1} \ominus \overrightarrow{0 w_2} := \overrightarrow{0 w}, \quad \text{where} \quad w = \frac{w_1 - w_2}{1 - w_1 \overline{w_2}} \quad (w_1, w_2 \in \mathbb{D}).$$

Then the hyperbolic PSO (HPSO) is defined by replacing $\cdot, +, -$ in Eq. (12) with their hyperbolic variants $\odot, \oplus, \ominus$. This algorithm can be applied directly in the single-pole case, i.e., when there is only one inverse pole, in other words $n = 1$ in Eq. (7) and its multiplicity is fixed.

### C. Hyperbolic PSO Algorithm for Multi-Pole Problems

The multi-pole problem is to determine the optimal inverse pole combination $\mathbf{a} = (a_0, \ldots, a_{n-1})^T \in \mathbb{D}^n$ with fixed multiplicities $\mathbf{m} = (m_0, \ldots, m_{n-1})^T \in \mathbb{N}_+^n$. For most of the heartbeats three poles are well separated according to the natural segmentation (P, T waves, QRS complex). Because of the localization property of basic rational functions the interference between terms with different poles is relatively small [44]. Therefore, we perform the optimization separately on each complex coordinate

$$\min_{a_i} r_2\big((a_0, \ldots, a_i, \ldots, a_{n-1})\big) \quad (i = 0, \ldots, n-1). \quad (15)$$

Then the solution of the multi-pole problem is reduced to successive applications of the single-pole optimization. It is a natural consequence of the hyperbolic model that the swarm cannot leave the unit circle during the algorithm. This makes the constraints $X_{\min}, X_{\max}$ used in the original Euclidean algorithm unnecessary.

In [45] we showed that the HPSO outperforms the well-known Nelder–Mead simplex algorithm in terms of reconstruction error and stability. The latter was proved to be important, especially when the MT system is used in classification problems [35]–[37].

## V. Generalized Rational Variable Projection With Optimization by Multi-Dimensional HPSO

In this section we give a non-trivial example for generalized variable projection inspired by a real applications, namely by signal compression. Keeping our eye on conditions (11) in Section III, we first choose a proper system. This will be the system of rational functions; the reasons behind it were provided in Section II-C4.

### A. Cost Function

The first term in Eq. (10), which measures how well the signal is represented by the projection, is set. Furthermore, it is natural to assume that the penalty term is strongly connected with the compression ratio in this case. According to this we specify the cost function as the linear combination of the approximation error and the reciprocal compression ratio (RCR) as follows:

$$\mathbf{f}_{cost}(\text{PRD}, \text{RCR}; \alpha) := \alpha \cdot \text{PRD} + (1 - \alpha) \cdot \text{RCR}, \quad (16)$$

where $\mathbf{f} \in \mathbb{R}^M$ is a discrete signal with $M$ samples, and

$$\text{PRD} := \|\mathbf{f} - \mathbf{P}^N_{\Theta(\mathbf{b})}\mathbf{f}\|_2 \cdot 100, \quad \text{RCR} := \frac{2 \cdot (n + N)}{M} \cdot 100.$$

Recall that $\mathbf{f}$ is of zero mean with unit variance, $n$ is the number of distinctive poles and $N$ is the sum of multiplicities (see Eq. (7)). The approximation error is computed as the usual percent root mean square difference (PRD). We point out that the second term containing RCR plays two roles. Namely, $n + N$ is obviously inversely proportional to the compression ratio on the algorithmic level. On the other hand, RCR can be viewed as the measure of complexity (dimension) of the system. The consequence of the penalty term is that the gain in PRD must reach a certain level in order to move to a higher dimension. The role of the regularization parameter $\alpha \in (0, 1]$ is to customize the method to different applications. The proper value of $\alpha$ for ECG compression is given in Section VI-C2. Since $\alpha$ is constant in a given problem, the cost function can be divided by $100 \cdot \alpha$ to obtain

$$\|\mathbf{f} - \mathbf{P}^N_{\Theta(\mathbf{b})}\mathbf{f}\|_2 + \frac{1 - \alpha}{\alpha} \cdot \frac{2 \cdot (n + N)}{M}. \quad (17)$$

The minimization of the cost function is designed to find the optimal pole configuration and the corresponding optimal pole positions. In order to do that we need to define an architecture space designed especially for the ECG signal compression problem.

TABLE I
ARCHITECTURE SPACE OF THE MDHPSO

| $d$ | $n+N$ | Conf. (**m**) | $d$ | $n+N$ | Conf. (**m**) | $d$ | $n+N$ | Conf. (**m**) |
|---|---|---|---|---|---|---|---|---|
| 1 | 8 | (2, 4) | 11 | 14 | (4, 8) | 21 | 19 | (4, 6, 6) |
| 2 | 9 | (8) | 12 | 15 | (4, 4, 4) | 22 | 19 | (4, 4, 8) |
| 3 | 10 | (4, 4) | 13 | 15 | (2, 2, 8) | 23 | 21 | (2, 8, 8) |
| 4 | 10 | (2, 6) | 14 | 15 | (2, 4, 6) | 24 | 21 | (4, 6, 8) |
| 5 | 11 | (2, 2, 4) | 15 | 16 | (6, 8) | 25 | 21 | (6, 6, 6) |
| 6 | 12 | (4, 6) | 16 | 17 | (4, 4, 6) | 26 | 23 | (4, 8, 8) |
| 7 | 12 | (2, 8) | 17 | 17 | (2, 6, 6) | 27 | 23 | (6, 6, 8) |
| 8 | 13 | (2, 4, 4) | 18 | 17 | (2, 4, 8) | 28 | 25 | (6, 8, 8) |
| 9 | 13 | (2, 2, 6) | 19 | 18 | (8, 8) | 29 | 27 | (8, 8, 8) |
| 10 | 14 | (6, 6) | 20 | 19 | (2, 6, 8) | 30 | 33 | (8, 14, 8) |

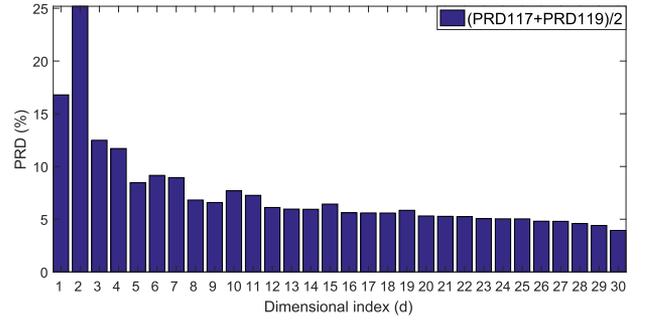

Fig. 1. Average PRDs of each pole configuration for the records 117, 119 of the PhysioNet MIT-BIH Arrhythmia dataset [8].

### B. Architecture Space

The pole configurations are described by the parameter vector **m** in Eq. (7). We note that the structure of the configuration set is rather complicated in the sense that the subspaces generated by different configurations are not comparable even if the numbers of the inverse poles along with the total dimensions of the subspaces, which are the sums of the multiplicities, are the same (see e.g. the cases (2, 6, 2) and (2, 4, 4)). This is a more complicated situation than the case of nested subspaces, which naturally induces a linear ordering and leads to the problem in Eq. (9). In the next step we will simplify the parameter set, i.e., we assign a virtual dimension $d$ to each parameter vector **m**, in such a way that it follows the system complexity $n + N$. Based on our experience we chose a set of 30 configurations that are worth considering in the ECG compression problem. Table I contains the pole configurations **m**, system complexity $n + N$ and the virtual dimension $d$. We note that there is however another expectation concerning dimensions. Namely, it is natural to expect that the increase of $d$ results in better PRD. Figs. 1 show that taking records 117 and 119 from the dataset [8] our ordering defined purely on the basis of system complexity $n + N$ behaves adequately in this respect. It is easy to see that by introducing the virtual dimension $d$, Eq. (17) can even formally be considered as an example for the generalized variable projection functional in Eq. (10) with $\Lambda(d) = \frac{1-\alpha}{\alpha} \cdot \frac{2 \cdot (n+N)}{M}$, where the connection between $d$ and $n + N$ is given in Table I.

In former works [32], [46], [47], it turned out that three inverse poles are sufficient for accurate representation of the heartbeats. These inverse poles are well separated and their multiplicities reflects the natural segmentation (P, T waves, QRS complex) of the heartbeat. Therefore, the inverse pole that corresponds to the QRS complex usually dominates the approximation, i.e., its multiplicity is higher than the others. The second and third most significant inverse poles correspond to the T and the P wave, respectively. Of course, in case of abnormal heartbeats some of these waves can be missing. These observations justify the pole configurations in Table I that we chose especially for ECG compression.

## C. Multi-Dimensional HPSO

After having defined the cost function, i.e., the generalized variable projection functional, we turn to the problem of optimization. To this order we take the so-called multi-dimensional (MD) PSO algorithm introduced by Kiranyaz et al. [6], and we adapt it to our case. MDPSO is a generalization of PSO, which along with the process of adaptation was established in Section IV. PSO based algorithms are constructed for static environments, but many practical problems change dynamically. This motivated the generalization of PSO to MDPSO, in which the dimensions are not fixed a priori. Then the optimization becomes a mixed integer nonlinear programming (MINLP) problem. The native structure of the swarm was extended by dimensional parameters. Thus, the particles can seek both positional and dimensional optima. The MDPSO was originally developed to evolve Artificial Neural Networks (ANN) for supervised learning [48], where the weights and bias of the network should be determined in order to minimize the classification error. We emphasize that since there is no penalty term concerning the structure of the ANNs, this optimization problem is still a variable projection method (see Section 3.2 in [48] and Section 1 in [9]). The original MDPSO algorithm [6] is the following

**Position updates:**

$$\mathbf{v}_k^{d_k} = c_1 r_1 \cdot \left(\widetilde{\mathbf{y}}_k^{d_k} - \mathbf{x}_k^{d_k}\right) + c_2 r_2 \cdot \left(\widehat{\mathbf{y}}^{d_k} - \mathbf{x}_k^{d_k}\right) + w \cdot \mathbf{v}_k^{d_k},$$
$$\mathbf{x}_k^{d_k} = \mathbf{x}_k^{d_k} + \mathbf{v}_k^{d_k}, \quad (18)$$

**Dimension updates:**

$$vd_k = \left[c_1 r_1 \cdot \left(\widetilde{d}_k - d_k\right) + c_2 r_2 \cdot \left(\widehat{d} - d_k\right) + vd_k\right],$$
$$d_k = d_k + vd_k, \quad (19)$$

where [.] is the integer rounding operator. The main changes compared to PSO given in Eq. (12) are the dimensional indices $d_k, \widetilde{d}_k, \widehat{d} \in \mathcal{I} = \{d_{\min}, \ldots, d_{\max}\}$, which denote the current, personal and global best dimensions, respectively. In this case, every particle has a certain position and velocity in each dimension. For instance, $\mathbf{x}_k^{d_k}$ denotes the position of the $k$th particle at the dimension $d_k \in \mathcal{I}$. The dimensions are kept within $\mathcal{I}$ by using so-called clamping operator. Note that in this algorithm the dimension parameter is a natural number assigned to every ANN structure.

Following the reasoning given in Section IV we adapt MDPSO to rational systems by replacing the arithmetic operations in the position update equations Eq. (18) by their hyperbolic variants. We call this modification MDHPSO and provide its pseudocode Alg. 1 in the Appendix. Recall that the vector **m** in Eq. (7) is the natural characterization of the complexity of the rational system. This was converted in the sequence $1, \ldots, 30$ in Table I. The real explanation of this conversion was to make our

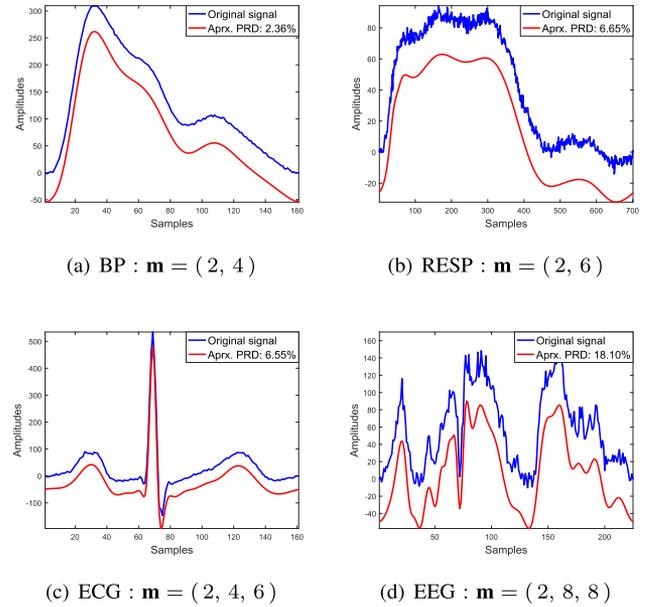

(a) BP : **m** = ( 2, 4 )  (b) RESP : **m** = ( 2, 6 )

(c) ECG : **m** = ( 2, 4, 6 )  (d) EEG : **m** = ( 2, 8, 8 )

Fig. 2. Approximation of biomedical signals with $\alpha = 0.5$ by using the first four channels of the record *slp02a* from the MIT-BIH/slpdb database [8].

optimization problem compatible with MDHPSO. In this way we made sure that our generalized rational variable projection constructed for ECG meets all the three conditions in (11).

Fig. 2 shows examples where the multi-dimensional rational variable model along with MDHPSO is applied to different types of signals such as blood pressure (BP), respiration (RESP), ECG and EEG. It is transparent that the method automatically adjusts the number of the inverse poles and the coefficients to the complexity of the signal. For this reason, only two inverse poles are used to represent the BP and RESP signals with 6 and 8 coefficients, and dimensional parameters $d = 1, 4$ respectively. In case of ECG and EEG signals the optimal dimensions of the architecture space are increased. The number of inverse poles changes from two to three, the number of coefficients are 12 and 18, and the dimensional parameters are $d = 14, 23$. We note that the solution produced by the MDHPSO can be refined by fast local methods such as Gauss-Newton algorithms staying within the optimal dimension.

## VI. ECG COMPRESSION

Biomedical monitoring of the human body is one of the most important tools for patient's diagnosis. It is for instance principal for proactive prevention of diseases. We note that long-term recordings such as ECG Holter-monitoring or 24 hours multi-channel electroencephalogram (EEG) recordings generate a large amount of data. This explains the need for compression in such cases. Of course the variable projection can also be used for various tasks other than compression including features extraction for classification, person identification, etc. Here we choose the task of compression of ECG to demonstrate the efficiency of our method.

A wide range of algorithms have been proposed in this field. They can be classified into three categories [49], namely,

parameter extraction algorithms, direct time-domain methods, and transform-domain techniques. Here we are considering the latter one, which can be interpreted as projections of the signal to low dimensional subspaces. Existing methods use sinusoids, wavelets, wavelet packets, Walsh functions, orthogonal polynomials, splines, principal components, etc. [13], [50]–[53]. For instance, Karczewicz and Gabbouj [12] proposed an algorithm that approximates the signal by linear combinations of B-splines with free knots. Although the algorithm is highly adaptive the computational cost is quite large due to the lack of orthogonality. In order to overcome this problem, orthogonal polynomials (in particular Hermite functions) were applied [13], [15], [16]. As a consequence of orthogonality, the coefficients of the Hermite representation can be easily calculated by using scalar products of the corresponding Hilbert space. Efficient implementations for both the continuous and the discrete cases are presented in [14], [17]. In contrast with the B-splines, Hermite-based compression schemes contain only one free parameter, namely the dilation of the base functions. Analogously, parametrized orthogonal wavelets can also be applied for signal compression. Then a wavelet decomposition with $L/2 - 1$ degrees-of-freedom is obtained, where $L$ is the length of the wavelet filter. Theoretically the number of free parameters is infinite, but for a large number of parameters the formulas become unmanageable. For this reason, in most of the signal processing applications the length of the filter is restricted to $L = 6$, which means only two degrees-of-freedom (see e.g. [24], [25], [54]). We note that only a few methods utilize rational functions [32], [46], [47], [55], [56], and even in those cases, the complexity of the models is fixed a priori. We will show that our approach is essentially different from them, and the proposed algorithm outperforms these methods.

### A. Preprocessing Stage

*1) Beat Detection/Normalization:* The compression method is based on successive evaluations of the MDHPSO algorithm on each heartbeat. The QRS complexes should be detected first to identify the heartbeats [57]. Then we follow [58] to get the segmentation. Namely, the original signal is cut at every 130th sample before each QRS peak. In the next step the linear correction is applied to avoid jumps at the endpoints:

$$f_i^* := f_i - \left(f_0 + i\frac{f_{M-1} - f_0}{M - 1}\right) \quad (0 \leq i < M),$$

where $\mathbf{f} = (f_0, f_1, \ldots, f_{M-1})^T$ is the discrete signal segment and $M$ is the number of samples. Then we apply normalization $\widetilde{\mathbf{f}} := \mathbf{f}^*/\|\mathbf{f}^*\|_2$. For the reconstruction of $\mathbf{f}$ the values $f_0, f_{M-1}$ and $\|\mathbf{f}^*\|_2$ should be stored as well.

*2) Hilbert Transform:* We will apply a rational transform, given in Section II-C4, for ECG records. Recall that the signal in Eq. (8) belongs to $H^2(\mathbb{D})$. This implies that the real function representing a heartbeat should be extended to a complex valued function in $H^2(\mathbb{D})$. It can be preformed by means of the well-known Hilbert transform. Therefore we will employ the discrete Hilbert transform $\mathcal{H}$ to $\widetilde{\mathbf{f}}$ to obtain $\widetilde{\mathbf{F}} := \widetilde{\mathbf{f}} + i\mathcal{H}\widetilde{\mathbf{f}}$.

TABLE II
DATA STRUCTURE OF THE COMPRESSED FILE

| Header | bits | Data of each beat | bits |
|---|---|---|---|
| architecture space | $30 \cdot 3 \cdot 4$ | best dimension | 6 |
| number of beats | 16 | $f_0, f_{M-1}, \|\mathbf{f}^*\|_2$ | 8, 8, 8 |
| mean of the ECG | 32 | poles: $\angle a_j, |a_j|$ | 4, 4 |
| std of the ECG | 32 | coeffs: $\angle c_j, |c_j|$ | 7, 7 |

### B. Compressing Stage

For the compression of $\widetilde{\mathbf{F}}$ obtained during preprocessing we will apply the multi-dimensional generalized rational variable projection method along with MDHPSO developed in Section V. Note that the original MDPSO was successfully applied in optimization problems related to dynamical environments [6]. From this point of view, the problem of ECG compression is similar due to the physiological behavior of the human heart. Although, the ECG signals are characterized by strong interbeat correlation, the segments are influenced by several factors including the respiration rate. Namely, the heart rate (HR) increases during inhalation and decreases during exhalation. In some cases only a few coefficients are needed for a good approximation. In other cases more coefficients are required to store the significant diagnostic information. Typical examples are the abnormal heartbeats, in which sudden changes are present.

*1) Basic Algorithm:* After the preprocessing stage, the MDHPSO algorithm and the corresponding rational projection are executed on each heartbeat. The result is an optimal inverse pole configuration, which is quantized and stored together with the related coefficients. The architecture space in Table I is also saved in the header of the compressed file. It serves as a look-up table for the pole configurations. The structure of the compressed file and the block diagram of the method can be seen in Table II and in Fig. 5.

*2) Aligned Algorithm:* As mentioned above, ECG signals have strong interbeat redundancy caused by cardiac cycles. Taking advantage of this phenomenon, compression results can be highly improved in certain situations. For this reason, we apply the average beat subtraction technique [59]. In this approach, the length of the heartbeats are equalized by zero-padding after segmentation and beat alignment. Then, the average of the first 30 beats in a record is approximated by using the pole configuration of the highest dimension of the architecture space. It provides an accurate representation of the mean cycle which is subtracted from all the segments of the ECG. Finally, the basic compression algorithm is applied on the residual signal. Although, the parameters of the average beat should also be stored in the header, higher accuracy/compression ratio is expected due to the interbeat correlation.

### C. Parameter Estimation

*1) Initial Swarm:* In order to speed up the convergence of the optimization, we use a starting estimate for the system parameters. Namely, we keep the optimal pole configurations of the previous segments in the initial swarm of the MDHPSO algorithm. More precisely, MDHPSO is initialized with a random swarm in which the first $a_{nb}$ particles contain the optimal pole configurations of the previous anb beats. anb is the average number of beats in a respiratory cycle computed by anb = [HR/brc], where brc denotes the number of breaths counted in a minute. In our implementation we set anb = 5.

TABLE III
PREDICTION RANGES OF PRDN AND WWPRD

| Measure | Quality groups | | | | |
|---|---|---|---|---|---|
| | Excellent (E) | V. Good (V) | Good (G) | Not Bad (N) | Bad (B) |
| PRDN | 0-4.33 | 4.33-7.8 | 7.8-11.59 | 11.59-22.5 | 22.5< |
| WWPRD | 0-7.4 | 7.4-15.45 | 15.45-25.18 | 25.18-37.4 | 37.4< |

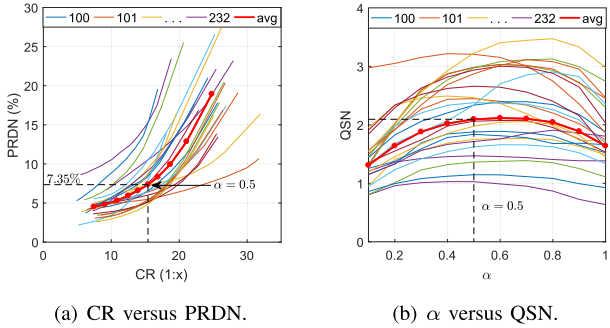

(a) CR versus PRDN.  (b) $\alpha$ versus QSN.

Fig. 3. Influence of the parameter $\alpha$ on the PRDN, CR, and QSN.

*2) Calibration of Parameter $\alpha$:* We investigated the influence of the regularization parameter $\alpha$ in the cost function $\mathbf{f}_{cost}$ on CR and PRD. Our goal was to achieve the highest CR by keeping PRD within the "very good range" according to Table III (cf. [60]). To this end, we took 24 records (see e.g. Table IV) of the MIT-BIH Arrhythmia database from Physionet [8]. Both the *basic* and the *aligned* versions of the proposed algorithm were applied on the signals by setting $\alpha = j/10, (j = 1, \ldots, 10)$. By Fig. 3 we conclude that 0.5 meets our requirements for these records. Indeed, we evaluated the normalized quality scores (QSN) in Eq. (21), which indicates that the optimal value of $\alpha$ is around 0.5. Although the performance is slightly better in terms of QSN for $\alpha = 0.6$, the corresponding PRD degrades to the "good" quality range. Therefore, we performed the tests for $\alpha = 0.5$. The results in Table IV show that this value is a good choice generally.

*3) Quantization:* The optimal inverse poles and the corresponding coefficients are linearly quantized and stored in the compressed file. In order to find the optimal number of bits for the representations, we use the records 117 and 119. These are extremely irregular ECGs, which are widely used for evaluating the performance of ECG compression algorithms (see e.g. [25], [58]). We need $k$ bits to represent the arguments $\angle a_j, \angle c_j$ and the absolute values $|a_j|, |c_j|$ if the quantization steps are $2\pi/2^k$ and $1/2^k$, respectively. Indeed, each beat is divided by its $\ell_2$ norm, so the energy of the signal is equal to 1. As a consequence of the orthogonality of the MT system and the Parseval's theorem, the constraint $|c_j| \leq 1$ holds. Moreover, $a_j \in \mathbb{D}$, so $1/2^k$ is a proper quantization step for both $|a_j|$ and $|c_j|$. In order to find the optimal quantization step, we executed the MDHPSO 100 times on the first four minutes of the records 117 and 119 for each quantization step. The average reconstruction errors

TABLE IV: Compression results of different variable projection methods.

| Records | Beat types | | Reconstruction error (PRDN %) | | | | | Diagnostic distortion (WWPRD %) | | | | | Compression ratio (CR 1 : X) | | | | | |
|---|---|---|---|---|---|---|---|---|---|---|---|---|---|---|---|---|---|---|
| | Normal | Abnormal | Basic | Aligned | B-spline | Hermite | AWT | AWPT | Basic | Aligned | B-spline | Hermite | AWT | AWPT | Basic | Aligned | B-spline | Hermite | AWT | AWPT |
| 100 | 2229 | 34 | 7.53 (G) | 8.27 (G) | 7.79 (V) | 7.35 (V) | 7.39 (V) | 7.01 (V) | 11.38 (V) | 12.35 (V) | 11.66 (V) | 10.64 (V) | 11.86 (V) | 21.70 (G) | 14.14 | 20.38 | 12.00 | 8.77 | 8.14 | 5.33 |
| 101 | 1853 | 5 | 6.78 (V) | 8.15 (G) | 7.03 (V) | 7.79 (V) | 6.62 (V) | 6.19 (V) | 11.16 (V) | 12.03 (V) | 11.36 (V) | 12.54 (V) | 12.33 (V) | 19.34 (G) | 16.58 | 20.59 | 12.74 | 10.68 | 8.74 | 5.96 |
| 102 | 99 | 2079 | 7.45 (V) | 8.98 (G) | 7.28 (V) | 21.87 (N) | 7.32 (V) | 5.86 (V) | 12.95 (V) | 15.74 (G) | 13.16 (V) | 38.14 (B) | 11.66 (V) | 19.01 (G) | 13.05 | 19.04 | 9.82 | 9.11 | 7.76 | 4.65 |
| 103 | 2074 | 2 | 5.36 (V) | 5.57 (V) | 5.44 (V) | 5.74 (V) | 5.25 (V) | 5.06 (V) | 9.33 (V) | 9.33 (V) | 9.27 (V) | 10.02 (V) | 10.77 (V) | 15.76 (G) | 15.98 | 21.43 | 12.17 | 9.56 | 8.31 | 5.63 |
| 104 | 163 | 2057 | 7.83 (G) | 8.76 (G) | 7.97 (G) | 18.50 (N) | 7.63 (V) | 7.17 (V) | 13.45 (V) | 15.48 (G) | 14.10 (V) | 37.18 (N) | 12.47 (V) | 16.91 (G) | 12.70 | 13.91 | 9.23 | 8.94 | 7.71 | 5.71 |
| 105 | 2517 | 46 | 7.25 (V) | 8.08 (G) | 7.87 (G) | 7.33 (V) | 7.24 (V) | 7.28 (V) | 16.27 (G) | 16.84 (G) | 17.02 (G) | 17.92 (G) | 18.49 (G) | 23.56 (G) | 14.94 | 19.19 | 11.37 | 7.79 | 8.51 | 6.90 |
| 106 | 1502 | 517 | 6.66 (V) | 6.86 (V) | 6.75 (V) | 7.92 (G) | 6.49 (V) | 6.15 (V) | 13.18 (V) | 13.80 (V) | 13.47 (V) | 15.01 (V) | 14.00 (V) | 19.03 (G) | 15.55 | 16.31 | 12.56 | 9.56 | 7.71 | 6.39 |
| 115 | 1944 | 0 | 5.15 (V) | 5.90 (V) | 5.03 (V) | 5.88 (V) | 5.05 (V) | 5.28 (V) | 9.02 (V) | 9.66 (V) | 8.86 (V) | 9.51 (V) | 10.23 (V) | 19.03 (G) | 15.37 | 23.19 | 12.40 | 10.21 | 8.41 | 5.54 |
| 117 | 1527 | 1 | 8.30 (G) | 8.91 (G) | 8.09 (G) | 9.87 (G) | 8.15 (G) | 8.26 (G) | 16.40 (G) | 16.97 (G) | 16.11 (G) | 18.25 (G) | 15.99 (G) | 20.99 (G) | 20.46 | 24.48 | 13.83 | 12.98 | 9.48 | 7.18 |
| 118 | 2159 | 110 | 8.38 (G) | 9.33 (G) | 8.63 (G) | 12.33 (N) | 8.23 (G) | 8.11 (G) | 14.79 (G) | 15.78 (G) | 15.08 (V) | 20.23 (G) | 17.32 (G) | 21.95 (G) | 14.37 | 19.79 | 10.28 | 8.75 | 7.49 | 6.04 |
| 119 | 1535 | 444 | 5.09 (V) | 5.29 (V) | 5.16 (V) | 6.28 (V) | 4.93 (V) | 4.54 (V) | 11.72 (V) | 12.68 (V) | 12.68 (V) | 15.68 (G) | 14.45 (V) | 25.60 (N) | 15.12 | 16.55 | 12.55 | 10.03 | 8.69 | 6.27 |
| 201 | 1630 | 327 | 7.06 (V) | 7.36 (V) | 7.37 (V) | 8.93 (G) | 6.87 (V) | 6.71 (V) | 13.38 (V) | 13.72 (V) | 14.21 (V) | 21.46 (G) | 15.00 (V) | 18.32 (G) | 16.80 | 19.46 | 14.95 | 10.39 | 9.99 | 7.23 |
| 202 | 2051 | 75 | 6.74 (V) | 7.16 (V) | 7.07 (V) | 6.81 (V) | 6.58 (V) | 6.53 (V) | 13.73 (V) | 14.19 (V) | 14.18 (V) | 20.07 (G) | 15.36 (V) | 20.96 (G) | 17.93 | 22.12 | 14.17 | 9.66 | 9.68 | 6.99 |
| 205 | 2561 | 85 | 7.86 (G) | 8.63 (G) | 8.16 (G) | 7.39 (V) | 7.73 (V) | 7.40 (V) | 12.06 (V) | 12.85 (V) | 13.17 (V) | 14.22 (V) | 13.56 (V) | 19.45 (G) | 14.44 | 19.70 | 11.20 | 7.56 | 7.72 | 5.30 |
| 207 | 1542 | 306 | 7.67 (V) | 8.33 (G) | 8.65 (G) | 7.11 (V) | 7.36 (V) | 7.10 (V) | 21.77 (G) | 22.80 (G) | 24.10 (G) | 22.54 (G) | 23.81 (G) | 37.18 (N) | 24.59 | 25.30 | 18.09 | 17.17 | 12.02 | 8.97 |
| 208 | 1581 | 1363 | 6.80 (V) | 7.30 (V) | 7.40 (V) | 7.03 (V) | 6.61 (V) | 6.39 (V) | 17.08 (G) | 17.08 (G) | 18.84 (G) | 25.03 (G) | 17.49 (G) | 22.90 (G) | 13.44 | 13.27 | 10.76 | 6.97 | 8.32 | 6.38 |
| 209 | 2611 | 384 | 12.39 (N) | 9.99 (G) | 13.81 (N) | 9.55 (G) | 12.03 (N) | 11.88 (N) | 14.01 (V) | 14.29 (V) | 19.05 (G) | 21.50 (G) | 18.64 (G) | 28.59 (N) | 12.74 | 16.02 | 12.83 | 6.93 | 8.41 | 5.98 |
| 212 | 2738 | 0 | 8.87 (G) | 10.40 (G) | 9.07 (G) | 10.28 (G) | 8.69 (G) | 8.47 (G) | 12.89 (V) | 14.91 (V) | 13.38 (V) | 14.90 (V) | 14.93 (V) | 18.95 (V) | 12.10 | 17.52 | 8.73 | 7.25 | 6.54 | 5.07 |
| 213 | 2630 | 609 | 5.62 (V) | 5.60 (V) | 6.12 (V) | 5.27 (V) | 5.49 (V) | 5.45 (V) | 12.37 (V) | 12.83 (V) | 13.45 (V) | 14.05 (V) | 18.62 (G) | 23.69 (G) | 13.17 | 15.92 | 11.29 | 6.18 | 7.56 | 5.76 |
| 214 | 1995 | 257 | 5.84 (V) | 5.88 (V) | 6.08 (V) | 6.23 (V) | 5.67 (V) | 5.44 (V) | 14.76 (V) | 14.91 (V) | 15.61 (G) | 21.14 (G) | 16.79 (G) | 23.96 (G) | 17.15 | 21.21 | 14.52 | 8.99 | 9.78 | 7.11 |
| 215 | 3183 | 168 | 10.65 (G) | 10.53 (G) | 11.76 (N) | 10.14 (G) | 10.45 (G) | 10.29 (G) | 14.81 (V) | 14.83 (V) | 16.77 (G) | 16.00 (G) | 17.33 (G) | 22.04 (G) | 12.22 | 15.06 | 10.12 | 6.03 | 7.39 | 5.52 |
| 217 | 244 | 1956 | 5.29 (V) | 5.92 (V) | 5.47 (V) | 9.98 (G) | 5.10 (V) | 4.94 (V) | 14.83 (V) | 16.08 (G) | 15.14 (V) | 25.50 (N) | 17.65 (G) | 23.66 (G) | 15.18 | 18.07 | 15.05 | 9.03 | 9.46 | 7.10 |
| 219 | 2074 | 71 | 4.67 (V) | 5.16 (V) | 4.61 (V) | 5.33 (V) | 4.54 (V) | 4.21 (E) | 11.53 (V) | 11.75 (V) | 11.48 (V) | 13.09 (V) | 14.12 (V) | 18.99 (G) | 15.08 | 20.37 | 15.05 | 9.26 | 8.90 | 6.35 |
| 232 | 396 | 1378 | 11.17 (G) | 12.15 (N) | 12.20 (N) | 16.35 (N) | 11.83 (N) | 11.86 (N) | 15.76 (G) | 16.90 (G) | 17.23 (G) | 25.09 (G) | 19.07 (G) | 25.53 (N) | 16.42 | 21.22 | 10.63 | 11.41 | 7.96 | 6.78 |
| Average | - | - | 7.35 (V) | 7.85 (G) | 7.70 (V) | 9.22 (G) | 7.22 (V) | 6.98 (V) | 13.69 (V) | 14.45 (V) | 14.56 (V) | 19.16 (G) | 15.50 (V) | 24.35 (G) | 15.40 | 19.17 | 12.28 | 9.31 | 8.57 | 6.26 |
| Std | - | - | 1.96 | 1.87 | 2.27 | 4.22 | 2.00 | 2.05 | 2.70 | 2.82 | 3.31 | 7.41 | 3.19 | 11.90 | 2.78 | 3.15 | 2.15 | 2.35 | 1.12 | 0.93 |

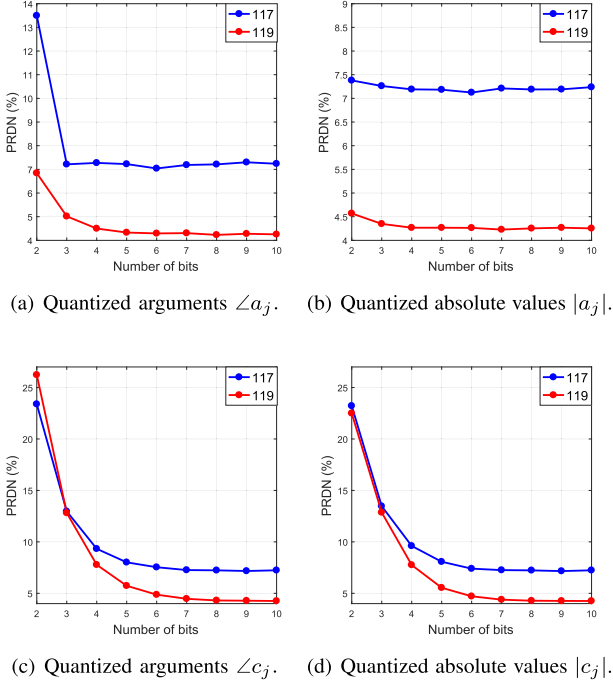

(a) Quantized arguments $\angle a_j$.    (b) Quantized absolute values $|a_j|$.

(c) Quantized arguments $\angle c_j$.    (d) Quantized absolute values $|c_j|$.

Fig. 4. Quantized arguments and absolute values of the inverse poles and the coefficients.

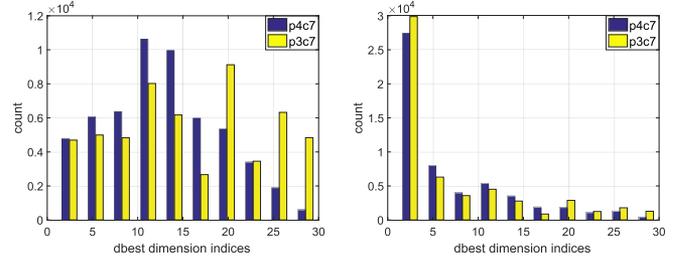

(a) Best dimensions (basic method).   (b) Best dimensions (aligned method).

Fig. 6. Comparison of the best dimensions for *p3c7* and *p4c7*.

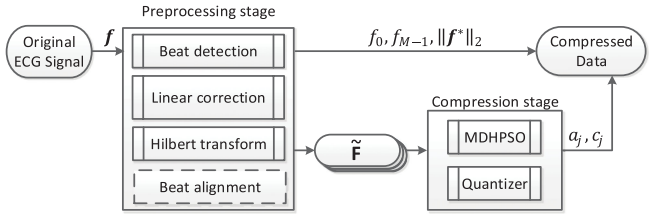

Fig. 5. Block diagram of the proposed compression algorithm.

of the runs for $k = 2, \ldots, 10$ is shown in Fig. 4. It is evident that 4 bits are enough to represent the angles and the absolute values of the inverse poles. Similarly, it is not worth to store the coefficients using more than 7 bits, since it does not improve the PRD significantly. The number of bits assigned to the quantities included in the compressed file are displayed in Table II.

We note that the quantization of the inverse poles is especially important, since it directly influences the optimization. In order to demonstrate this effect, we examined the compression performance on the same 24 records as in Table IV by using two quantization configurations: *p3c7* and *p4c7*. The abbreviaton *pxcy* denotes the number of bits we used to store both the absolute values and angles of the poles (p) and the coefficients (c), e.g. *p4c7* denotes the optimal setup. As was to be expected, the sub-optimal setup *p3c7* yielded worse reconstruction error (PRD) and compression ratio (CR) compared to *p4c7*. The reason is twofold. On one hand, storing the parameters on a better resolution decreases the PRD. On the other hand, in case of *p3c7*, it turned out that the low resolution of the inverse poles were compensated by an increased number of coefficients. In the optimal setup *p4c7*, the optimization choose less complex rational function systems to minimize the cost function which means less coefficients to store, i.e., better CR. This phenomenon can be seen in Fig. 6, where we displayed the histograms of the best dimension indices at the final iteration of the optimization. The figure shows that the optimization for *p4c7* terminated more frequently in lower dimensions than its counterpart *p3c7*.

## VII. EXPERIMENTS

For comparison tests we used the MIT/BIH arrhythmia database [8], which has been widely used for testing ECG compression techniques. The dataset contains 48 half-hour long ECG recordings, which was digitized to 11-bit resolution. We note that in many papers the tests were performed for only a small portion (couple of records, minutes) of the whole set. Here we take those 24 records suggested in [61], and compress the first channel of the entire signal. It results in a total of 12 hours long raw ECG data which was used to compare the proposed algorithm and those described in Section II-C. We note that we implemented the recent versions of the corresponding algorithms. Namely, the Hermite expansions of the QRS complexes were calculated by discrete orthogonal polynomials [14] and the mother wavelet parametrization was applied along with wavelet packets optimization [24]. In the latter case, the wavelet domain was represented via the modified run-length coding recommended in [25].

Although the reconstruction error is usually expressed in terms of PRD, we cannot ignore the distortion measures designed especially for ECG signals. Unfortunately, only a few distortion criteria are available in ECG compression for performance evaluations. One of them is the so-called *weighted diagnostic distortion* (WDD) measure [62]. On one hand, it correlates very well with mean opinion scores (MOS) of the clinical experts. On the other hand, there is no standard code for computing the WDD. Besides, the measure is unstable due to the requirement of accurate classification for characteristic features of the ECG signal. For this reason, we will use another diagnostic distortion measure called the wavelet-based weighted (WW) PRD [60]. In that case, the signal is decomposed into five sub-bands which are weighted regarding their cardiological significance. Let us denote the coefficient vectors of the $j$th wavelet level of the original and the reconstructed signals with zero mean by $\mathbf{c}_j$ and $\tilde{\tilde{\mathbf{c}}}_j$, respectively. Then the WWPRD can be

defined as follows:

$$\text{WWPRD} = 100 \times \sum_{j=0}^{5} \mathbf{w}_j \cdot \frac{\left\| \mathbf{c}_j - \tilde{\tilde{\mathbf{c}}}_j \right\|_2}{\left\| \mathbf{c}_j \right\|_2}. \quad (20)$$

The sequence of weights are $\mathbf{w} = \frac{6}{27}, \frac{9}{27}, \frac{7}{27}, \frac{3}{27}, \frac{1}{27}, \frac{1}{27}$ which were heuristically assigned to each sub-band emphasizing their diagnostic significance. We note that both the PRD and the WWPRD measures provide five quality groups corresponding to different ranges of the values, which are summarized in Table III (see Table VII in [60]). According to [63], the PRD should be computed for zero mean signals. Then it is not affected by the mean of the signal. In order to distinguish the two types of PRD we will use the following notations:

$$\text{PRDN} = 100 \times \frac{\left\| \mathbf{f} - \tilde{\tilde{\mathbf{f}}} \right\|_2}{\left\| \mathbf{f} - \text{mean}(\mathbf{f}) \right\|_2}, \quad \text{PRD} = 100 \times \frac{\left\| \mathbf{f} - \tilde{\tilde{\mathbf{f}}} \right\|_2}{\left\| \mathbf{f} \right\|_2},$$

where $\tilde{\tilde{\mathbf{f}}}$ is the reconstructed signal, and PRDN stands for the normalized PRD, i.e., PRD of a signal with zero mean.

We performed the comparison tests in two stages. First, the proposed algorithm, both the basic and the aligned variants, were executed by using the optimal parameters given in Section VI-C. Second, we took the four other variable projection methods discussed in Section II-C: B-splines [12], Hermite functions [13], [14], wavelets [25], wavelet packets [24], [64], [65]. In order to make a fair comparison, we executed these four algorithms iteratively till their PRDNs became close to those of our basic method for each record. We increased the number of coefficients gradually, and we stopped the iteration if the PRDN of our method (the values in the fourth column of Table IV) or the maximum number of coefficients was reached. As a consequence, sometimes the PRDN was not even close to that of our method (see e.g. the Hermite representation for the record 102). One can see the results of the experiment in Table IV, where CR denotes the ratio of the number of bits of the original and the compressed files. Note that we used the normalized definition of the PRDN which is independent of the mean of the signal. In addition, we calculated the PRDN for each beat and presented their average for the records in Table IV.

## VIII. DISCUSSION

First, we compare the basic and the aligned variations of the proposed compression algorithm. As it was shown in [59], the regularity of ECG cycles facilitates the application of beat subtraction techniques. This complies with the significant improvement that can be seen for more than half of the records: 100, 103, 105, 115, 201, 202, 205, 209, 213, 214, 215, 217, 219, 232. Namely, in these cases the mean of the differences in PRDNs and WWPRDs are 0.41% and 0.62%, and the compression ratio is 1.38 times better in the aligned case, on the average. These results agree with those in [59], especially for the records 103, 201, 202, 213, 214, 215, 217, 219 (see e.g. Fig. 8 in [59]). The standard deviations of PRDN, WWPRD and CR for the aligned and the basic algorithms are close to each other and quite low. This indicates the stability of the algorithms. Moreover, according to Table III, the average PRDN and WWPRD of the reconstructed ECG signals fall within the range of very good and good quality for the basic and the aligned variation of the proposed method. On the other hand the aligned method is not always preferable, see e.g. the records 104, 207, 208 in Table IV. In these cases, the average beat estimation is poor due to the high presence of abnormal beats and irregular cardiac cycles. According to [59], the compression performance of the simple average beat subtraction technique highly depends on the similarity between the estimated average beat and the beats to be compressed. Therefore, a big improvement can be achieved for those recordings that present one type of beats in a regular rhythm. In order to analyze the performance of the aligned algorithm, we computed the average Pearson correlation coefficient $\rho_{x,y}$ between the estimated average beat and the beats to be compressed. As we discussed, the rhythm is also important, hence we calculated the standard deviation $\sigma_{RR}$ of the R to R peak distances. Since the compression performance is proportional to $1/\sigma_{RR}$ and $\rho_{x,y}$, we analyzed these measures along with the normalized quality score differences:

$$\Delta \text{QSN} = \text{QSN}_{\text{aligned}} - \text{QSN}_{\text{basic}},$$

where QSN is defined in Eq. (21). Fig. 7 shows that the average beat subtraction improved the compression performance for almost every record, except 104, 207 and 208. In case of 207 and 208, both $1/\sigma_{RR}$ and $\rho_{x,y}$ are low, i.e., the average beat estimations are bad and the rhythms are irregular. In Fig. 8, we displayed the first 30 beats of the record 207. The set comprises two completely different beat types, thus their simple average will be a bad estimate of any of them. Record 104 is another example, which contains a lot of paced (No 1373) and fusion (No 664) beats. The difference between these two beat types can be seen in Fig. 9(a). Also, Fig. 9(b) shows the first 30 beats of the record 104. Although the estimated average beat is quite similar to the paced beats, it is still very different from the fused ones. This is why the resulting QSN is lower for the aligned method, despite of the high $1/\sigma_{\text{RR}}$ and $\rho_{x,y}$ values. We note that there are other more sophisticated beat subtraction techniques including average beat code books. Based on the results on regular ECG records, we expect that the proposed method can be further improved by using these techniques. This will be a part of future work.

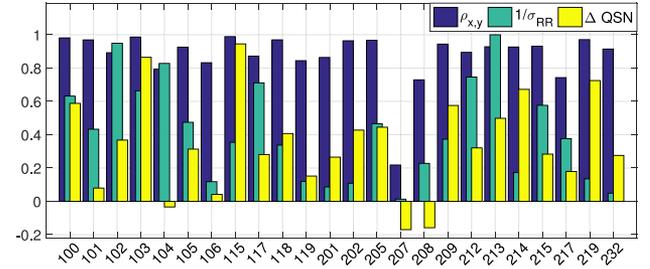

Fig. 7. Analysis of the compression performance of the basic and the aligned algorithm.

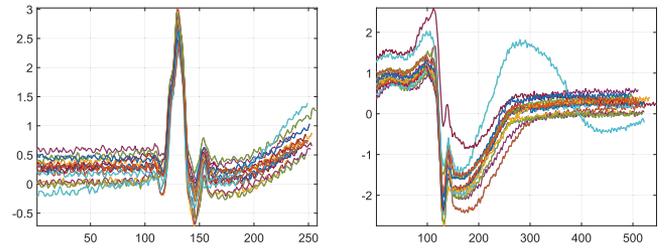

(a) Right bundle branch block beats. (b) Premature ventricular contractions.

Fig. 8. The first 30 beats in the record 207.

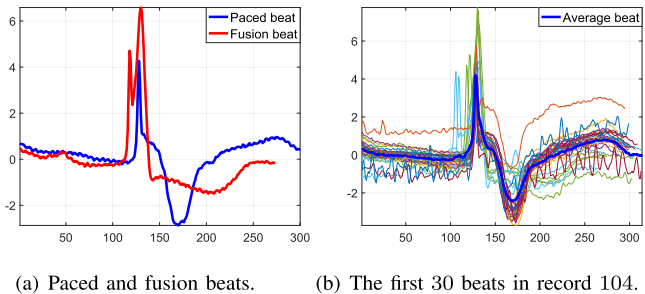

(a) Paced and fusion beats.  (b) The first 30 beats in record 104.

Fig. 9. Beat types of the record 104.

TABLE V
COMPARISON OF PROPOSED METHODS WITH THE STATE-OF-THE-ART

| Tested records: same as in Tab. IV | | | | Tested records: 100, 101, 102, 103, 115, 117, 118, 119, 213, 232 | | | |
|---|---|---|---|---|---|---|---|
| Method | PRD | CR | QS | Method | PRDN | CR | QSN |
| Tan et al. [56] | 1.42 | 35.52 | 33.80 | Agulhari et al. [67] | 4.06 | 9.51 | 2.34 |
| Ma et al. [55] | 1.04 | 25.64 | 30.96 | Rajoub [68] | 4.30 | 5.79 | 1.35 |
| Fira et al. [61] | 1.17 | 18.27 | 20.71 | Benzid et al. [69] | 4.24 | 8.21 | 1.94 |
| Lee et al. [70] | 1.90 | 21.35 | 14.11 | Lu et al. [71] | 3.71 | 6.60 | 1.78 |
| Yildirim et al. [72] | 2.70 | 32.25 | 13.76 | Bilgin et al. [73] | 4.13 | 9.53 | 2.32 |
| Proposed Basic | 0.45 | 15.40 | 37.99 | Proposed Basic | 7.08 | 15.47 | 2.30 |
| Proposed Aligned | 0.47 | 19.17 | 44.33 | Proposed Aligned | 7.82 | 20.26 | 2.75 |
| Tested records: 117, 119 | | | | Tested records: 100, 101, 102, 103, 105, 117, 201, 202, 205, 207, 232 | | | |
| Method | PRD | CR | QS | Method | PRD | CR | QS |
| Abo-Zahhad et al. [25] | 1.90 | 24.00 | 12.87 | Fathi et al. [74] | 3.13 | 28.59 | 10.07 |
| Proposed Basic | 0.44 | 17.79 | 41.57 | Proposed Basic | 0.35 | 16.85 | 49.75 |
| Proposed Aligned | 0.46 | 20.51 | 45.70 | Proposed Aligned | 0.38 | 21.17 | 57.72 |

Now, we compare the proposed method with the other four ECG models discussed in Section II-C. We note that also these algorithms can be modified by beat subtraction techniques. Therefore, in order to make a fair comparison, we exclude the aligned method from the analysis. Moreover, as mentioned in Section VII, every method was adjusted to have PRDNs similar to the basic method. This is reflected in the "Reconstruction error" section of Table IV. Comparing the WWPRDs we conclude that the basic method is superior to the others both in terms of average and standard deviation. The same holds for most of the individual records as well. It is worth mentioning that the averages of both PRDN and WWPRD fall in the very good quality category. The best values for the standard deviations indicate the reliability of getting high quality compressed signals with CR 1:15. It is evident that the highest CR, which was our ultimate motivation, was achieved by our algorithm. The CR values for the others, except for the B-spline case, are significantly lower. In terms of WWPRD and CR the B-spline method turned to be the second best. We note that it has much higher computational cost compared to the basic method due to the iterative knot removal algorithm and the lack of orthogonality.

The variable projection via Hermite functions is one of the least effective methods in sense of PRDN, and WWPRD. In fact, the desired level of PRDN could not be reached, within the predefined low compression ratio in several cases. This is due to the segment based representation. We get a good approximation of a wave/spike if the main peak is close to the middle of the segment. On the other hand, if the predictions of the endpoints of the P, QRS, T waves are not good enough, then the approximation error is higher. The three free dilation parameters available cannot compensate the effect of bad segmentation.

We note that the optimization of AWT with respect to the two free parameters always terminated near the values $a_1 = 1.3598$, $a_2 = -0.7821$. The average differences are less than $7.7 \cdot 10^{-3}$ and $1.1 \cdot 10^{-2}$, respectively. These are the parameters of the Daubechies wavelet db3. This means that the advantage of the adaptivity of this method is negligible in ECG signals. This complies with the experience in [24]. The idea behind the optimal behavior of the db3 wavelets for ECG signals lie in the facts that they have three vanishing moments (see e.g. [23], [66]), and ECG curves are quite smooth signals. The latter one explains the tendency of using B-splines, orthogonal polynomials in this field.

The optimization of the adaptive wavelet packet transform (AWPT) is carried out in two steps. Following [24], we selected the best basis from the wavelet packet tree with depth 7 according to the Shannon entropy. Then, we changed the parameters of the mother wavelet using Eq. (6). As shown in Table IV, the CR values of AWPT are worse than those of AWT. This is due to the fact that the optimal parameters were again close to the wavelet basis db3. In contrast to the simple AWT, we have to store the best wavelet basis as well, which explains the low CR values. Our observation again coincides with the results of [24]. In that work, the same adaptive wavelets and wavelet packets were applied as in Section II-C for compressing ECG and EMG signals. Although they achieved a high improvement for EMG signals, the results on ECG curves were very close to the db3 wavelet basis. Hence, they concluded that the gain in performance is strongly depended on the signal type. We call the attention to the fact that the diagnostic distortion (WWPRD), which was not examined in previous works like [24], [64], [65], is significantly higher for AWPT than for AWT.

We also compare the proposed algorithm to state-of-the-art methods that are based on various approaches including wavelets [25], [67], [68], wavelet packets [74], wavelet encoding [69], [71], discrete cosine transform [70], image compression [73], and deep neural networks [72]. In order to evaluate the performance of these methods the so-called quality scores were introduced in [61]:

$$\text{QS} = \text{CR}/\text{PRD}, \quad \text{QSN} = \text{CR}/\text{PRDN}. \quad (21)$$

The higher the QS or QSN, the better the performance. In Table V, we summarized the results for those records that are common in the cited papers and in our work. The results show that the *aligned* variation of the *basic* algorithm outperforms all the others. In addition, even the *basic* procedure is very close to the competing methods. Particularly, the QS is very high for [55], [56], which are also utilizing rational functions. However, in those works, the complexity of the model, i.e., the number of inverse poles $n$, and the subspace dimension $N$ were fixed a priori, while these parameters are found automatically in our algorithm. It is also worth mentioning that most of the competing methods in Table V apply entropy based lossless compression techniques as a final step. These algorithms can be utilized for the proposed method as well, and it would further improve our results.

Although the proposed generalized rational variable projection algorithm is quite complex numerically the time complexity is manageable. Let $N_{it}$ stand for the number of iteration of the MDHPSO, and recall the notations for the swarm size $S$ and for the number of dimensions $|\mathcal{I}|$. At the initialization step we need to evaluate the cost function in Eq. (16) at each particle and for each dimension. It is followed by $S$ number of function evaluation for each iteration. Therefore the overall number of function evaluation is equal to $S \cdot |\mathcal{I}| + S \cdot N_{it}$. In our experiments the algorithms were implemented in MATLAB. An Intel(R) Core(TM) i7-6700 @ 3.40 GHz CPU was used. The size of the swarm and the number of iteration were set to 30 and 20 in the optimization, respectively. Then the average execution time of a 30 minutes long ECG record is about 23 minutes for both the basic and the aligned algorithms. The only competitive algorithm in terms of WWPRD and CR is the B-spline method, for which the average execution time is about 91 minutes. Note that the B-spline method utilizes a knot removal algorithm in which the initial number of knots is proportional to the number of samples in a beat. Hence, high sampling rate increases the execution time. In contrary, the computational complexity of our method depends on the number of function evaluations and the number of iterations only, which are predefined manually. In addition, one can reduce the number of relevant inverse pole configurations in Table I based a priori information of specific class of signals. The execution time can be further decreased by applying parallel implementations of the PSO and the MDHPSO algorithms, but these are beyond the scope of this work.

## IX. CONCLUSION

In the first part of this paper we considered a mathematical model called variable projection and showed that several known signal processing methods can be understood as special case of that. Then we studied the rational function systems in this framework. We generalized the rational variable projection model by adding the model complexity as free parameters. For the optimization of the system parameters we developed the multi-dimensional hyperbolic variant of the well-known PSO algorithm. Based on our experience we believe that our method can be effectively used for various problems in signal processing including data fitting, feature extraction, segmentation, etc. In this work the problem of compressing ECG signals was chosen for demonstrating its efficiency. It turned out that the generalized rational variable projection algorithm outperforms the previously known ones. The MATLAB implementations of all the algorithms included in this paper and the test results are available at the website [75], [76]. We emphasize that the proposed algorithm is of general nature. Namely it is flexible and can be adjusted to different types of signals like EEG [35].

## APPENDIX
## ALGORITHMS

**Algorithm 1:** Multi-Dimensional Hyperbolic Particle Swarm Optimization (interactive code is available at [76]).

1: **function** MDHPSO $S$, $N_{it}$, $\alpha$
2:    $w_0 := 0.2$, $w_N := 0.8$    ▷ inertia weights
3:    $c_1 := 1.5$, $c_2 := 2$    ▷ learning factors
4:    **for all** $d \leftarrow d_{min}, \ldots, d_{max}$ **do**
5:      **for all** $k \leftarrow 1, \ldots, S$ **do**
6:        Randomize $\mathbf{x}_k^d$, $\mathbf{v}_k^d$    ▷ initialize particles
7:        Initialize $\widetilde{\mathbf{y}}_k^d := \mathbf{x}_k^d$
8:        Randomize $d_k$, $vd_k$    ▷ initialize dimensions
9:        Initialize $\widetilde{d}_k := d_k$
10:      **end for**
11:      Initialize $\text{gbest}^d := \arg\min_{1 \leq k \leq S} \rho(\mathbf{x}_k^d, d)$
12:      Initialize $\widehat{\mathbf{y}}^d := \mathbf{x}_{\text{gbest}^d}^d$
13:    **end for**
14:    Initialize $\widehat{d} := \arg\min_{d_{min} \leq d \leq d_{max}} \rho(\widehat{\mathbf{y}}^d, d)$
15:    **for** $\ell \leftarrow 1, \ldots, N_{it}$ **do**
16:      $w := w_N - \ell(w_N - w_0)/N_{it}$    ▷ update weights
17:      **for** $k \leftarrow 1, \ldots, S$ **do**
18:        **if** $\rho(\mathbf{x}_k^{d_k}, d_k) < \rho(\widetilde{\mathbf{y}}_k^{d_k}, d_k)$ **then**
19:          $\widetilde{\mathbf{y}}_k^{d_k} := \mathbf{x}_k^{d_k}$    ▷ update local best positions
20:          **if** $\rho(\mathbf{x}_k^{d_k}, d_k) < \rho(\widetilde{\mathbf{y}}_k^{\widetilde{d}_k}, \widetilde{d}_k)$ **then**
21:            $\widetilde{d}_k := d_k$    ▷ update the local best dim
22:          **end if**
23:        **end if**
24:        $\rho_{d_k} := \min\left\{\rho(\widehat{\mathbf{y}}^{d_k}, d_k), \min_{1 \leq i \leq S, i \neq k} \rho(\mathbf{x}_i^{d_k}, d_k)\right\}$
25:        **if** $\rho(\mathbf{x}_k^{d_k}, d_k) < \rho_{d_k}$ **then**
26:          $\text{gbest}^{d_k} := k$    ▷ update global best pos
27:          $\widehat{\mathbf{y}}^{d_k} := \mathbf{x}_k^{d_k}$
28:          **if** $\rho(\mathbf{x}_k^{d_k}, d_k) < \rho(\widehat{\mathbf{y}}^{\widehat{d}}, \widehat{d})$ **then**
29:            $\widehat{d} := d_k$    ▷ update the global best dim
30:          **end if**
31:        **end if**
32:      **end for**
33:      Update all $\mathbf{x}_k^{d_k}$, $\mathbf{v}_k^{d_k}$ according to Eq. (18)
34:      Update all $d_k$, $vd_k$ according to Eq. (19)
35:    **end for**
36:    **return** $\mathbf{x}_{\text{gbest}^{\widehat{d}}}^{\widehat{d}}$
37: **end Function**

## REFERENCES

[1] S. G. Mallat and Z. Zhang, "Matching pursuits with time-frequency dictionaries," *IEEE Trans. Signal Process.*, vol. 41, no. 12, pp. 3397–3415, Dec. 1993.

[2] S. S. Chen, D. L. Donoho, and M. A. Saunders, "Atomic decomposition by basis pursuit," *SIAM Rev.*, vol. 43, no. 1, pp. 129–159, 2001.

[3] R. R. Coifman, Y. Meyer, and M. V. Wickerhauser, "Wavelet analysis and signal processing," in *Wavelets and Their Applications*, M. B. Ruskai *et al.*, Eds., Boston, MA, USA: Jones and Bartlett, 1992, pp. 153–178.

[4] P. Stoica and Y. Selen, "Model-order selection: A review of information criterion rules," *IEEE Signal Process. Mag.*, vol. 21, no. 4, pp. 36–47, Jul. 2004.

[5] G. H. Golub and V. Pereyra, "The differentiation of pseudo-inverses and nonlinear least squares problems whose variables separate," *SIAM J. Numer. Anal.*, vol. 10, pp. 413–432, 1973.

[6] S. Kiranyaz, J. Pulkkinen, A. Yildirim, and M. Gabbouj, "Multi-dimensional particle swarm optimization in dynamic environments," *Expert Syst. Appl.*, vol. 38, no. 3, pp. 2212–2223, 2011.

[7] S. Kiranyaz, T. Ince, and M. Gabbouj, "Multidimensional particle swarm optimization for machine learning and pattern recognition," in *Adaptation, Learning, and Optimization*, vol. 15. Berlin, Germany: Springer, 2014.

[8] A. L. Goldberger et al., "PhysioBank, physioToolkit, and physioNet: Components of a new research resource for complex physiologic signals," *Circulation*, vol. 101, no. 23, pp. 215–220, 2000.

[9] G. H. Golub and V. Pereyra, "Separable nonlinear least squares: The variable projection method and its applications," *Inverse Problems*, vol. 19, no. 2, pp. R1–R26, 2003.

[10] V. Pereyra and G. Scherer, Eds., *Exponential Data Fitting and Its Applications*. Oak Park, IL, USA: Bentham Science Publishers, 2010.

[11] D. L. B. Jupp, "Approximation to data by splines with free knots," *SIAM J. Numer. Anal.*, vol. 15, no. 2, pp. 328–343, 1978.

[12] M. Karczewicz and M. Gabbouj, "ECG data compression by spline approximation," *Signal Process.*, vol. 59, pp. 43–59, 1997.

[13] R. Jane, S. Olmos, P. Laguna, and P. Caminal, "Adaptive Hermite models for ECG data compression: Performance and evaluation with automatic wave detection," in *Proc. Int. Conf. Comput. Cardiology*, 1993, pp. 389–392.

[14] A. Sandryhaila, S. Saba, M. Püschel, and J. Kovacevic, "Efficient compression of QRS complexes using Hermite expansion," *IEEE Trans. Signal Process.*, vol. 60, no. 2, pp. 947–955, Feb. 2012.

[15] H. Haraldsson, L. Edenbrandt, and M. Ohlsson, "Detecting acute myocardial infarction in the 12-lead ECG using hermite expansions and neural networks," *Artif. Intell. Med.*, vol. 32, pp. 127–136, 2004.

[16] M. Lagerholm, C. Peterson, G. Braccini, L. Edenbrandth, and L. Sörnmo, "Clustering ECG complexes using hermite functions and self-organizing maps," *IEEE Trans. Biomed. Eng.*, vol. 47, no. 7, pp. 838–717, Jul. 2000.

[17] L. Sörnmo, P. L. Börjesson, M. E. Nygards, and O. Pahlm, "A method for evaluation of QRS shape features using a mathematical model for the ECG," *IEEE Trans. Biomed. Eng.*, vol. 28, no. 10, pp. 713–717, Oct. 1981.

[18] G. Georgiev, I. Valova, N. Gueorguieva, and L. Lei, "QRS complex detector implementing orthonormal functions," *Proc. Comput. Sci.*, vol. 12, pp. 426–431, 2012.

[19] G. Szegő, *Orthogonal Polynomials*, 3rd ed. New York, NY, USA: AMS Colloquium Publications, 1967.

[20] W. Gautschi, *Orthogonal Polynomials, Computation and Approximation* (Numerical Mathematics and Scientific Computation). Oxford, U.K.: Oxford Univ. Press, 2004.

[21] S. Mallat, *A Wavelet Tour of Signal Processing: The Sparse Way*, 3rd ed. Burlington, NJ, USA: Academic, 2008.

[22] P. S. Addison, *The Illustrated Wavelet Transform Handbook – Introductory Theory and Applications in Science, Engineering, Medicine and Finance*. Bristol, U.K.: Institute of Physics Publishing, 2002.

[23] C. S. Burrus, A. R. Gopinath, and H. Guo, *Introduction to Wavelets and Wavelet Transforms: A Primer*, 1st ed. Englewood Cliffs, NJ, USA: Prentice-Hall, 1997, ch. 5.

[24] L. Brechet, M. F. Lucas, C. Doncarli, and D. Farina, "Compression of biomedical signals with mother wavelet optimization and best-basis wavelet packet selection," *IEEE Trans. Biomed. Eng.*, vol. 54, no. 12, pp. 2186–2192, Dec. 2007.

[25] M. Abo-Zahhad, A. F. Al-Ajlouni, S. M. Ahmed, and R. J. Schilling, "A new algorithm for the compression of ECG signals based on mother wavelet parameterization and best-threshold level selection," *Digit. Signal Process.*, vol. 23, no. 3, pp. 1002–1011, 2013.

[26] A. Bultheel, P. Gonzalez-Vera, E. Hendriksen, and O. Njåstad, *Orthogonal Rational Functions*. Cambridge, U.K.: Cambridge Univ. Press, 2009.

[27] F. Malmquist, "Sur la détermination d'une classe fonctions analytiques par leurs dans un ensemble donné de points," *Compte Rendus Six. Cong. Math. Scand.*, Kopenhagen, Denmark, pp. 253–259, 1925.

[28] S. Takenaka, "On the orthogonal functions and a new formula of interpolations," *Japanese J. Math.*, vol. 2, pp. 129–145, 1925.

[29] P. Kovács and L. Lócsi, "RAIT: The rational approximation and interpolation toolbox for MATLAB," in *Proc. 35th Int. Conf. Telecomm. Signal Process.*, 2012, pp. 671–677.

[30] P. Kovács and V. Vad, "Fast computing of non-uniform sampling positions for real signals," in *Proc. 15th Int. Symp. Symbolic Numeric Algorithms Sci. Comput.*, 2013, pp. 146–150.

[31] S. Fridli and F. Schipp, "Biorthogonal systems to rational functions," *Annales Univ. Sci. Budapest., Sect. Comp*, vol. 35, pp. 95–105, 2011.

[32] S. Fridli, L. Lócsi, and F. Schipp, "Rational function system in ECG processing," in *Computer Aided Systems Theory, Part I*, vol. 6927, R. Moreno-Díaz et al., Eds., Berlin, Germany: Springer, 2012, pp. 88–95.

[33] P. S. C. Heuberger, P. M. J. Van den Hof, and B. Wahlberg, *Modelling and Identification With Rational Orthogonal Basis Functions*, London, U.K.: Springer, 2005.

[34] P. Kovács, K. Samiee, and M. Gabbouj, "On application of rational discrete short time Fourier transform in epileptic seizure classification," in *Proc. 39th IEEE Int. Conf. Acoust., Speech, Signal Process.*, 2014, pp. 5839–5843.

[35] K. Samiee, P. Kovács, and M. Gabbouj, "Epileptic seizure classification of EEG time-series using rational discrete short time Fourier transform," *IEEE Trans. Biomed. Eng.*, vol. 62, no. 2, pp. 541–552, Feb. 2015.

[36] K. Samiee, P. Kovács, and M. Gabbouj, "Epileptic seizure detection in long-term EEG records using sparse rational decomposition and local Gabor binary patterns feature extraction," *Knowl.-Based Syst.*, vol. 118, pp. 228–240, 2017.

[37] K. Samiee, P. Kovács, S. Kiranyaz, M. Gabbouj, and T. Saramäki, "Sleep stage classification using sparse rational decomposition of single channel EEG records," in *Proc. 23rd Eur. Signal Process. Conf.*, 2015, pp. 1905–1909.

[38] J. Kennedy and R. C. Eberhart, "Particle swarm optimization," in *Proc. IEEE Int. Conf. Neural Netw.*, 1995, vol. 4, pp. 1942–1948.

[39] Y. Shi and R. Eberhart, "A modified particle swarm optimizer," in *Proc. IEEE Int. Conf. Evol. Comput., IEEE World Congr. Comput. Intell.*, 1998, pp. 69–73.

[40] I. C. Trelea, "The particle swarm optimization algorithm: Convergence analysis and parameter selection," *Inf. Process. Lett.*, vol. 85, no. 6, pp. 317–325, 2003.

[41] F. Van den Bergh and A. P. Engelbrecht, "A study of particle swarm optimization particle trajectories," *Inf. Process. Lett.*, vol. 176, no. 8, pp. 937–971, 2006.

[42] F. Schipp, "Hyperbolic wavelets," in *Topics in Mathematical Analysis and Applications* (Optimization and Its Applications), vol. 94, T. Rassias and L. Tóth, Eds., Berlin, Germany: Springer, 2014, pp. 633–657.

[43] A. A. Ungar, *Analytic Hyperbolic Geometry: Mathematical Foundations and Applications*. Hackensack, NJ, USA: World Scientific Publishing, 2005.

[44] P. Kovács, "Rational variable projection methods in ECG signal processing," in *Computer Aided Systems Theory, Part II*, vol. 10672, R. Moreno-Díaz et al., Eds. Berlin, Germany: Springer, 2017, pp. 196–203.

[45] P. Kovács, S. Kiranyaz, and M. Gabbouj, "Hyperbolic particle swarm optimization with application in rational identification," in *Proc. 21st Eur. Signal Process. Conf.*, 2013, pp. 1–5.

[46] L. Lócsi and P. Kovács, "Processing ECG signals using rational function systems," in *Proc. 7th IEEE Int. Symp. Med. Meas. Appl.*, 2012, pp. 1–5.

[47] S. Fridli, P. Kovács, L. Lócsi, and F. Schipp, "Rational modeling of multi-lead QRS complexes in ECG signals," *Annales Univ. Sci. Budapest., Sect. Comp*, vol. 37, pp. 145–155, 2012.

[48] S. Kiranyaz, T. Ince, A. Yildirim, and M. Gabbouj, "Evolutionary artificial neural networks by multi-dimensional particle swarm optimization," *Neural Netw.*, vol. 22, no. 10, pp. 1448–1462, 2009.

[49] S. M. S. Jalaleddine, C. G. Hutchens, R. D. Strattan, and W. A. Coberly, "ECG data compression techniques—A unified approach," *IEEE Trans. Biomed. Eng.*, vol. 37, no. 4, pp. 329–343, Apr. 1990.

[50] H. Lee and K. M. Buckley, "ECG data compression using cut and align beats approach and 2-D transform," *IEEE Trans. Biomed. Eng.*, vol. 46, no. 5, pp. 556–564, May 1999.

[51] P. S. Addison, "Wavelet transforms and the ECG: A review," *Physiological Meas.*, vol. 26, no. 5, pp. 155–199, 2005.

[52] E. Berti, F. Chiaraluce, N. E. Evans, and J. J. McKee, "Reduction of Walsh-transformed electrocardiograms by double logarithmic coding," *IEEE Trans. Biomed. Eng.*, vol. 47, no. 11, pp. 1543–1547, Nov. 2000.

[53] F. Castells, P. Laguna, L. Sörnmo, A. Bollmann, and J. M. Roig, "Principal component analysis in ECG signal processing," *EURASIP J. Adv. Signal Process.*, vol. 2007, pp. 1–21, 2007.

[54] A. Maitrot, M.-F. Lucas, C. Doncarli, and D. Farina, "Signal-dependent wavelets for electromyogram classification," *Med. Biol. Eng. Comput.*, vol. 43, no. 4, pp. 487–492, 2005.

[55] J. Ma, T. Zhang, and M. Dong, "A novel ECG data compression method using adaptive Fourier decomposition with security guarantee in e-health applications," *IEEE J. Biomed. Health Informat.*, vol. 19, no. 3, pp. 986–994, May 2015.

[56] C. Tan, L. Zhang, and H. Wu, "A novel Blaschke unwinding adaptive Fourier decomposition based signal compression algorithm with application on ECG signals," *IEEE J. Biomed. Health Informat.*, vol. 23, no. 2, pp. 672–682, Mar. 2019.

[57] J. Pan and W. J. Tompkins, "A real-time QRS detection algorithm," *IEEE Trans. Biomed. Eng.*, vol. 32, no. 3, pp. 230–236, Mar. 1985.

[58] H. H. Chou, Y. J. Chen, Y. C. Shiau, and T. S. Kuo, "An effective and efficient compression algorithm for ECG signals with irregular periods," *IEEE Trans. Biomed. Eng.*, vol. 53, no. 6, pp. 1198–1205, Jun. 2006.

[59] P. S. Hamilton and W. J. Tompkins, "Compression of the ambulatory ECG by average beat subtraction and residual differencing," *IEEE Trans. Biomed. Eng.*, vol. 38, no. 3, pp. 253–259, Mar. 1991.

[60] A. S. Al-Fahoum, "Quality assessment of ECG compression techniques using a wavelet-based diagnostic measure," *IEEE Trans. Inf. Technol. Biomedicine*, vol. 10, no. 1, pp. 182–191, Jan. 2006.

[61] C. M. Fira and L. Goras, "An ECG signals compression method and its validation using NNs," *IEEE Trans. Biomed. Eng.*, vol. 55, no. 4, pp. 1319–1326, Apr. 2008.

[62] Y. Zigel, A. Cohen, and A. Katz, "The weighted diagnostic distortion (WDD) measure for ECG signal compression," *IEEE Trans. Biomed. Eng.*, vol. 47, no. 11, pp. 1424–1430, Nov. 2000.

[63] A. Alshamali and A. S. Al-Fahoum, "Comments on 'an efficient coding algorithm for the compression of ECG signals using the wavelet transform'," *IEEE Trans. Biomed. Eng.*, vol. 50, no. 8, pp. 1034–1037, Aug. 2003.

[64] M. L. Hilton, "Wavelet and wavelet packet compression of electrocardiograms," *IEEE Trans. Biomed. Eng.*, vol. 44, no. 5, pp. 394–402, May 1997.

[65] M. Blanco-Velasco, F. Cruz-Roldán, J. I. Godino-Llorente, and E. K. Barner, "Wavelet packets feasibility study for the design of an ECG compressor," *IEEE Trans. Biomed. Eng.*, vol. 54, no. 4, pp. 766–769, Apr. 2007.

[66] I. Daubechies, *Ten Lectures on Wavelets*, 1st ed. Philadelphia, PA, USA: SIAM, 1992.

[67] C. M. Agulhari, I. S. Bonatti, and P. L. D. Peres, "An adaptive run length encoding method for the compression of electrocardiograms," *Med. Eng. Phys.*, vol. 35, pp. 145–153, 2013.

[68] B. A. Rajoub, "An efficient coding algorithm for the compression of ECG signals using the wavelet transform," *IEEE Trans. Biomed. Eng.*, vol. 49, no. 4, pp. 355–362, Apr. 2002.

[69] R. Benzid, F. Marir, and N-E. Boughechal, "Electrocardiogram compression method based on the adaptive wavelet coefficients quantization combined to a modified two-role encoder," *IEEE Signal Process. Lett.*, vol. 14, no. 6, pp. 373–376, Jun. 2007.

[70] S. Lee, J. Kim, and M. Lee, "A real-time ECG data compression and transmission algorithm for an e-health device," *IEEE Trans. Biomed. Eng.*, vol. 58, no. 9, pp. 2448–2455, Sep. 2011.

[71] Z. Lu, D. Y. Kim, and W. A. Pearlman, "Wavelet compression of ECG signals by the set partitioning in hierarchical trees algorithm," *IEEE Trans. Biomed. Eng.*, vol. 47, no. 7, pp. 849–856, Jul. 2000.

[72] O. Yildirim, R. S. Tan, and U. R. Acharya, "An efficient compression of ECG signals using deep convolutional autoencoders," *Cogn. Syst. Res.*, vol. 52, pp. 198–211, 2018.

[73] A. Bilgin, M. W. Marcellin, and M. I. Altbach, "Compression of electrocardiogram signals using JPEG2000," *IEEE Trans. Consum. Electron.*, vol. 49, no. 4, pp. 833–840, Nov. 2003.

[74] A. Fathi and F. Faraji-Kheirabadi, "ECG compression method based on adaptive quantization of main wavelet packet subbands," *Signal, Image Video Process.*, vol. 10, no. 8, pp. 1433–1440, 2016.

[75] P. Kovács, "ECG compression algorithms," 2017. [Online]. Available: http://numanal.inf.elte.hu/~kovi/

[76] P. Kovács, S. Fridli, and F. Schipp, "Generalized rational variable projection with application in ECG compression," 2017. [Online]. Available: https://codeocean.com/capsule/8330505/tree/v1.

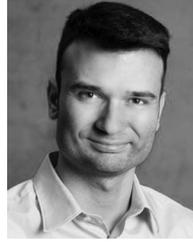

**Péter Kovács** received the B.Sc., M.Sc., and Ph.D. degrees in computer science all from the Eötvös Loránd University (ELTE), Budapest, Hungary, in 2008, 2010, and 2016, respectively. His research interests cover the topics of optimization, biomedical signal modeling, analysis and implementation of numerical algorithms. During his Ph.D. studies, he spent five months as a Visiting Researcher with the Department of Signal Processing, Tampere University of Technology, Finland. After receiving his Ph.D. degree in 2016, he was promoted to an Assistant Professor with the Department of Numerical Analysis, ELTE. Since March 2018, he has been a Postdoc with the Institute of Signal Processing, Johannes Kepler University Linz, Linz, Austria. He was the recipient of the Farkas Gyula Prize in applied mathematics (János Bolyai Mathematical Society) in 2016.

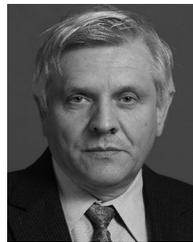

**Sándor Fridli** received the Ph.D. degree from Eötvös Loránd University (ELTE), Budapest, Hungary, in 1983 and the Doctor of Science degree from the Hungarian Academy of Sciences, Budapest, Hungary. He is a Professor and the Head of Department of Numerical Analysis, Faculty of Informatics, ELTE. In the academic years 1989/1991 and 1995/1996, he was a Visiting Professor with the University of Tennessee, Knoxville, USA, and in 2000/2001 with the University of Colorado, Colorado Springs, USA. He was invited to give colloquium talks at several universities including the University of British Columbia, Vancouver, Canada, University of Missouri, Columbia, USA, Reinisch-Westfälische Technische Hochschule, Aachen, Germany, and the Indian Society of Industrial and Applied Mathematics, ISIAM. He is a Founder Member of the Hungarian Service Network for Mathematics in Industry and Innovations, and a member of the Scientific Board of it. In 2018, the Szent-Györgyi Albert Prize of the Government of Hungary was awarded to him. His main research fields are applied harmonic analysis, approximation theory, and signal processing.

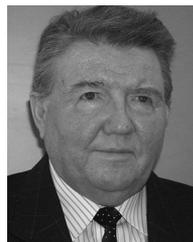

**Ferenc Schipp** is a Professor Emeritus with the Eötvös Loránd University (ELTE), Budapest, Hungary. Formerly, he was the Head and a Professor of the Numerical Analysis Department, ELTE for 20 years. His main research interests are harmonic analysis and its applications in signal processing, and in system and control theories. Some focal points of his research are dyadic analysis, martingale theory, wavelet theory, rational transforms. His book (with coauthors) about Walsh–Fourier series *Walsh Series: An Introduction to Dyadic Harmonic Analysis* (Adam Hilger) was published in 1990, and is the fundamental work in dyadic analysis ever since. He was invited to several institutes and universities worldwide. He is a Doctor et Professor Honoris Causa of the ELTE (2012), and of the University of Pécs (2017). In 2008, the Széchenyi prize was awarded to him by the Government of Hungary. In 2016, he received the Eötvös József Prize of the Hungarian Academy of Sciences.